\documentclass{article}

\usepackage{arxiv}

\usepackage[utf8]{inputenc} 
\usepackage[T1]{fontenc}    
\usepackage{hyperref}       
\usepackage{url}            
\usepackage{booktabs}       
\usepackage{amsmath, amsfonts}       
\usepackage{nicefrac}       
\usepackage{microtype}      
\usepackage{lipsum}		
\usepackage{graphicx,xcolor}
\usepackage{natbib}
\usepackage{doi}

\usepackage{caption}  
\usepackage{subcaption}
\usepackage{longtable}
\usepackage{lscape}
\usepackage{float}

\title{Searching for long faint astronomical high energy transients: a data driven approach}

\author{Riccardo Crupi \\
	DMIF\\
	Università di Udine\\
	Via delle Scienze 206, Udine 33100 \\
	\texttt{crupi.riccardo@spes.uniud.it} \\
	\And
	Giuseppe Dilillo \\
	Istituto di Astrofisica e Planetologia Spaziali\\
	INAF\\
	via del Fosso del Cavaliere 100, Roma 00113 \\
	\texttt{giuseppe.dilillo@inaf.it} \\
        \And
	Kester Ward \\
	STOR-i Doctoral Training Centre \\
        Lancaster University \\
        Lancaster, UK  \\
        \texttt{k.ward4@lancaster.ac.uk} \\
	\And
	Elisabetta Bissaldi \\
	Dipartimento Interateneo di Fisica \\
        Politecnico di Bari \\
        via E. Orabona 4, Bari 70125 \\ \\
 	Sezione di Bari \\
        Istituto Nazionale di Fisica Nucleare \\
        via E. Orabona 4, Bari 70125 \\
        \texttt{elisabetta.bissaldi@ba.infn.it} \\
	\And
	Fabrizio Fiore \\
	Osservatorio Astronomico di Trieste \\
        INAF \\
	via Tiepolo 11, Trieste 34143\\
	\texttt{fabrizio.fiore@inaf.it} \\
	\And
	Andrea Vacchi\\
        DMIF\\
	Università di Udine\\
	Via delle Scienze 206, Udine 33100 \\ \\
	Sezione di Trieste \\
        Istituto Nazionale di Fisica Nucleare \\
	via Padriciano 99, Trieste 34149\\
	\texttt{andrea.vacchi@uniud.it} \\
}

\renewcommand{\shorttitle}{Searching for long transients: a data driven approach}

\hypersetup{
pdftitle={Searching for long faint astronomical high energy transients: a data driven approach},
pdfsubject={astro-ph.HE, cs.LG, stat.CO},
pdfauthor={Riccardo Crupi, Giuseppe Dilillo, Elisabetta Bissaldi, Fabrizio Fiore, Andrea Vacchi},
pdfkeywords={Gamma-Ray Burst, Deep Learning, Trigger Algorithm, Background Estimation},
}

\begin{document}
\maketitle

\begin{abstract}
HERMES (High Energy Rapid Modular Ensemble of Satellites) pathfinder is an in-orbit demonstration consisting of a constellation of six 3U nano-satellites hosting simple but innovative detectors for the monitoring of cosmic high-energy transients. The main objective of HERMES Pathfinder is to prove that accurate position of high-energy cosmic transients can be obtained using miniaturized hardware. The transient position is obtained by studying the delay time of arrival of the signal to different detectors hosted by nano-satellites on low Earth orbits. To this purpose, particular attention is placed on optimizing the time accuracy, with the goal of reaching an overall accuracy of a fraction of a micro-second. In this context, we need to develop novel tools to fully exploit the future scientific data output of HERMES Pathfinder. 
In this paper, we introduce a new framework to assess the background count rate of a space-born, high energy detector; a key step towards the identification of faint astrophysical transients.  
We employ a Neural Network (NN) to estimate the background lightcurves on different timescales. Subsequently, we employ a fast change-point and anomaly detection technique to isolate observation segments where statistically significant excesses in the observed count rate relative to the background estimate exist. We test the new software on archival data from the NASA Fermi Gamma-ray Burst Monitor (GBM), which has a collecting area and background level of the same order of magnitude to those of HERMES Pathfinder. The neural network performances are discussed and analyzed over period of both high and low solar activity. We were able to confirm events in the Fermi/GBM catalog, both Solar Flares and Gamma-Ray Bursts (GRBs), and found events, not present in Fermi/GBM database, that could be attributed to Solar Flares, Terrestrial Gamma-ray Flashes, GRBs, Galactic X-ray flash. Seven of these are selected and analyzed further, providing an estimate of localisation and a tentative classification.
\end{abstract}

\section{Introduction}\label{sec1}

Gamma-Ray Bursts (GRBs) originate in extraordinarily energetic explosions taking place in distant galaxies. They appear as irregular pulses of X and $\gamma$-ray radiation in detectors of today work-horse satellites such as SWIFT, INTEGRAL, Fermi, and Agile. The typical distribution of GRBs duration is bimodal; `long bursts', lasting longer than 2s, are associated with black hole formation in collapsars, while `short bursts', lasting less than 2s, are associated to mergers of binary neutron stars \cite{woosley1993gamma, berger2014short}.

Present instrumentation dedicated to GRBs and cosmic transients has been launched during the 2010s. There is no guarantee that it will continue to operate beyond the mid-2020s. For this reason, several proposals to NASA and ESA have been already submitted to select the successors of these instruments. The High Energy Rapid Modular Ensemble of Satellites (HERMES) concept is to develop a constellation of nano-satellites to study high-energy transients \cite{fiore2020hermes,fiore2021}, thus providing a fast-track and affordable solution bridging the gap between current X-ray monitors and the next generation. A technological and scientific pathfinder (HERMES-TP, funded by ASI and HERMES-SP funded by the European Commission, HERMES pathfinder hereafter) is in preparation to prove the concept, that is the capability to detect and localize GRBs with miniaturized instrumentation hosted by nano-satellites. The first six HERMES Pathfinder spacecrafts are expected to be launched in low-Earth, near-equatorial orbit during 2024. A seventh payload unit identical to those hosted by HERMES Pathfinder is also hosted by the Australian SpiRIT satellite, to be launched during 2023. The HERMES Pathfinder and SpIRIT payload is a small yet innovative ``siswich" detector providing broad-band energy coverage (few keV - $1$~MeV) and very good temporal resolution (a few hundreds ns) \cite{fuschino2019hermes, evangelista2020scientific,fiore2022,evangelista2022}.

GRBs manifest as transient increases in the count rates of detectors. The activity of these phenomena appear as unexpected, and not explainable in terms of background or any other known sources. Any automated procedure for detecting GRBs is generally concerned with searching the time series of the observations for statistically significant excesses in photon counts, relative to a reference background estimate in the absence of $\gamma$/X-ray GRB related events. The on-orbit physical background observed by GRB monitor experiments is determined by factors inherent to the highly dynamical near-Earth radiation environment, to the spacecraft geographic position and attitude, as well as the spacecraft geometry, and the detector's pointing, design and response. 
Given the difficulty intrinsic to a real-time modelling of the expected scientific background, algorithms dedicated to the `online' search of GRBs often resort to extrapolate the background from recent observations. For example, the trigger algorithms running on-board NASA Fermi/GBM assess a background estimate from an average of the photon count rate observed over the previous $17$~s excluding the most recent $4$~s of observations \cite{meegan2009fermi}; similar moving average approaches were used by Compton-BATSE \cite{paciesas1999fourth} and BeppoSAX-GRBM \cite{feroci1997flight}.

In `offline' analysis, archival data are searched for GRB events that the online and on-board algorithms may have missed. Examples of this approach can be found in \cite{kommers1999faint}, which uses the BATSE catalog, or in \cite{kocevski2018analysis} and \cite{hui2017finding} where they search for faint, short GRBs at times compatible with known gravitational wave events.
In \cite{biltzinger2020physical} for example, an estimate is assessed starting from detailed models of the background expected for GBM, such as the detector response, the cosmic $\gamma$-ray background, the solar activity, the geomagnetic environment, the Earth albedo and the visibility of X and $\gamma$ point sources. 
The background description so achieved has been shown to reproduce very well the observations of Fermi/GBM and could potentially allow for the identification of otherwise hard to detect GRBs such as long-weak events with slow raising times. However, having been specifically tailored for the observations of Fermi/GBM, this technique is not immediately applicable to other experiments. 
In \cite{sadeh2019deep} a Recurrent Neural Network (RNN \cite{lecun2015deep}) is used to predict the background and, on top of it, classify or detect anomalies in the observations of a count-rate detector. To recognize a GRB event, this RNN is trained onto existing catalogues of burst observations. We believe such an approach could inherit the detection biases of standard strategies for GRB detection, ultimately leading to missing events which already defied previous searches.

In Section \ref{sec_method} we introduce our approach to estimate the scientific background of a gamma-ray burst monitor experiment using a Neural Network (NN). In particular, we employ a Feed Forward Neural Network (\cite{bishop1995neural}) to estimate the count-rates expected from background sources over the 12 NaI detectors of Fermi/GBM, in different energy bands and at regular time intervals. Our model is designed to learn the dynamics of the background over a timescale of months, enabling the detection of long-GRBs or even ultralong-GRBs \cite{gendre2019can}, as shown in an example in Section \ref{benchmark}.  Moreover, employing a robust loss function in the training phase, we are able to deal with outliers in count-rate observations, such as transients due to astronomical events or brief period of detector inactivity \ref{solarmaxmin}.
The choice of applying our framework to archival data from Fermi was motivated by the facts that (1) the HERMES Pathfinder spacecrafts are expected to be launched in a low inclination orbit with altitude $500-550$ km, an orbit where the background and its variations are expected to be smaller than those of Fermi/GBM \cite{meegan2009fermi}; and (2) the Fermi/GBM and HERMES Pathfinder detectors both rely onto scintillators and have similar effective areas \cite{bissaldi2009ground,campana2020hermes, dilillo2022space} resulting in background count-rates of the same order of magnitude. 
To estimate the background observed by Fermi/GBM, we leverage on a large ensemble of information, including features both intrinsic to the satellite and its orbital setting such as the satellite attitude and geographic location in time, the Sun visibility and so on. This idea is consistent with \cite{fitzpatrick2012background}, which describes a method that estimates the background at the period of interest by using rates from adjacent days when the satellite has similar geographical footprint. 
To retrieve these information's we use the Fermi/GBM Data Tools \cite{GbmDataTools} software package, an Application Programming Interface (API) allowing to download, analyse and visualise GBM data.
Being completely data-driven, we believe our approach to be in principle applicable to any GRB monitor experiment for which a similar dataset is available.

The background estimates produced by the NN are compared with the observations by mean of an efficient change-point detection technique called FOCuS-Poisson \cite{ward2022poisson}, aiming at the automatic identification of statistically significant astrophysical transients. We tested the combination of the NN background estimates and FOCuS-Poisson trigger on real Fermi/GBM data. We were able to confirm known events, but we also find events with no counterpart in the Fermi/GBM trigger catalog\footnote{https://heasarc.gsfc.nasa.gov/W3Browse/fermi/fermigtrig.html} \citep{von2020fourth}, yet with features resembling astronomical transients such as GRBs and solar flares and other galactic high-energy sources.

The paper is organised as follows. In Section \ref{sec_method} we present the background estimation in a supervised Machine Learning settings, the architecture of the NN and the FOCuS-Poisson change-point detection technique. In Section \ref{sec_data} we describe the data used and the pre-processing steps to build the dataset. In Section \ref{sec_results} we report the performance of the NN estimator and the result of the application of the trigger algorithm. A comparison between the background estimated in a period of solar maxima and in a solar minima is described in the Appendix \ref{solarmaxmin}.
In Section \ref{sec_analysis} we discuss the results of our search for undiscovered astrophysical transients. We identify 110 events with no counterpart in GBM trigger catalog over a period of about 9 months. We report on a subset of seven events providing lightcurves, localization and classification. 
Finally, in Section \ref{sec:conclusion} we draw our conclusions and discuss future prospects.  

\section{Methodology}\label{sec_method}
The background assessment problem is expressed as a supervised Machine Learning estimator, with the variables inherent to the satellite and its orbital position as inputs and the count-rate observed by each detector in three different energy bands as outputs.
The background estimates so obtained are compared against the actual observations using FOCuS-Poisson. 
The significance of the excess in the count observations relative to the background model is quantified in units of standard deviations and recorded as a time series. Finally, these records are searched for intervals in the observation where the excess significance exceeds a threshold over one or more detector-energy band combinations.

\subsection{Background estimation}\label{sec_background}
\begin{figure}
    \centering
    \includegraphics[width=1\textwidth]{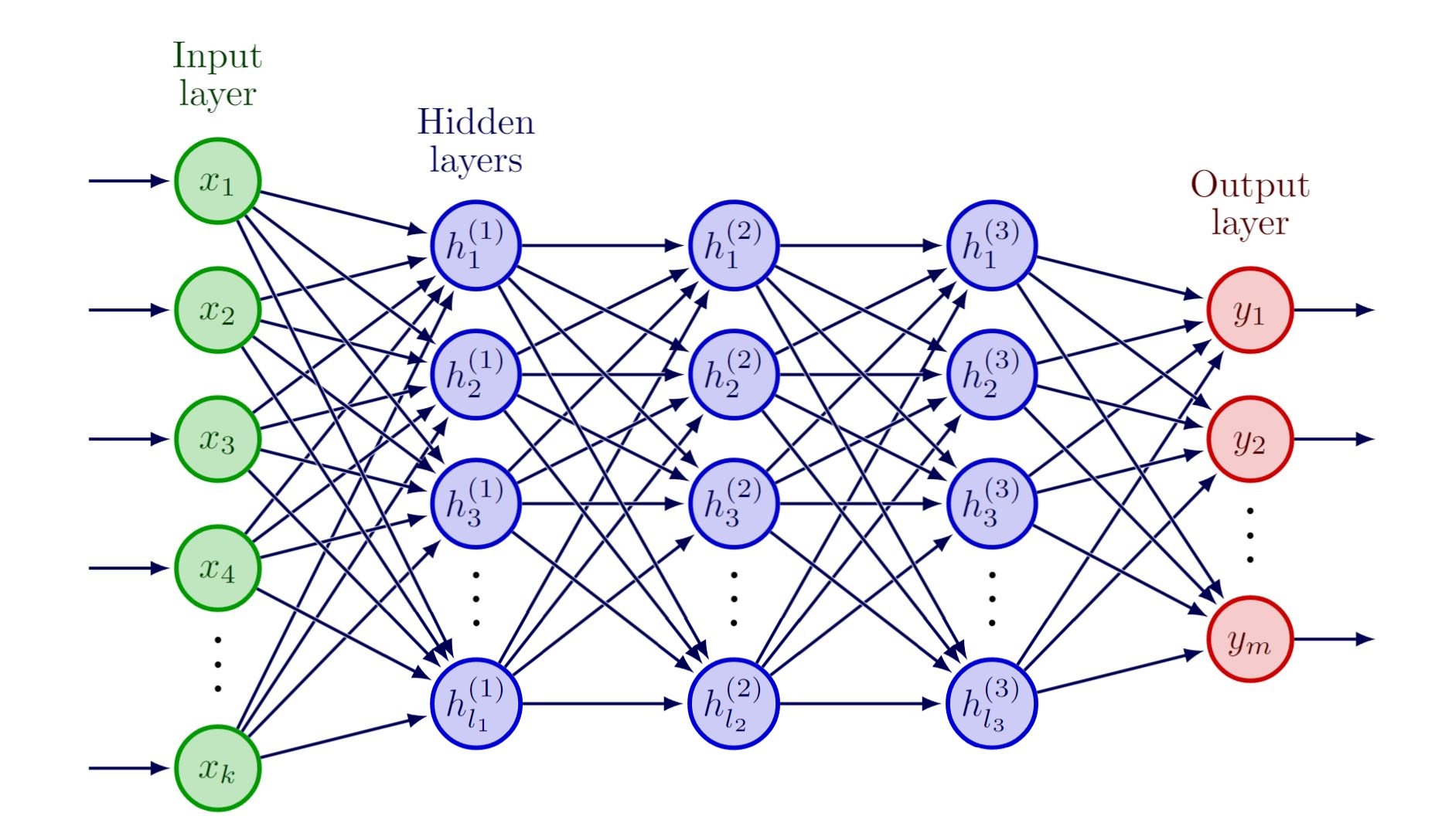}
    \caption{The architecture of a feed forward neural network. The input has dimension 60. The first two hidden layer have 2048 neurons, the third 1024. The output layer has dimension 36.}
    \label{fig:nn}
\end{figure}

 We define $X$ as the input variables, see $col\_sat\_pos$ and $col\_det\_pos$ in Section~\ref{sec_data}, and $Y$ as the output variables, see $col\_range$ in Section~\ref{sec_data}. 
We suppose that a function $f(X)$ exists which predict $Y$ given $X$, that is the solution that minimize $L(f(x), Y)$ ($\text{argmin}_f  L(f(x), Y)$) where $L$ is the loss function that quantify the error in the predictions. The model's goal is to estimate a quantity $F(x)$ such that $f(x) \approx F(x)$ \cite{hastie2009elements}. Here we are dealing with a multi-output regression: $F: X \in \mathbb{R}^{k} \longrightarrow Y \in \mathbb{R}^{m}$, where $k$ is the number of features into the model and $m$ the number of outputs.\\
The model employed is a feed forward neural network with 3 hidden dense layers (Figure~\ref{fig:nn}). Each hidden layer is followed by a batch normalization layer \cite{ioffe2015batch} and a dropout layer \cite{JMLR:v15:srivastava14a}. The NN is implemented in Tensorflow \cite{tensorflow2015-whitepaper}. 
The input layer has dimension $k=60$. Each of the first two hidden layers is composed of $2048$ neurons, while the third hidden layer hosts $1024$ neurons. The last (output) layer has $m=36$ neurons. Each of the output neurons is associated with a particular detector-energy combination. The probability parameter for the drouputs is $0.02$. The optimizer used is Nadam \cite{Ruder16} with learning rate $\eta$ varying accordingly to Equation \ref{lr_decay}, $\beta_1 = 0.9$, $\beta_2 = 0.99$ and $\epsilon = 10^{-7}$.
\begin{equation}\label{lr_decay}
  \eta =
    \begin{cases}
      0.01 & \text{if epoch $<4$}\\
      0.0016 & \text{if $4\ge$ epoch $<12$}\\
      0.0004 & \text{if epoch $\ge 12$}
    \end{cases}       
\end{equation}
We run the fitting for 64 epochs with a batch size of 2048.

In a pre-processing step, the input training dataset is standardised and filtered. Data filtering takes place in two steps in which the following data subsets are removed:
\begin{itemize}
    \item data collected while Fermi is transiting through the high radiation environment of the South Atlantic Anomaly (SAA).
    \item data acquired at times in which an event of the Fermi/GBM trigger catalog occurred.
\end{itemize}
This latter choice isn't strictly necessary, yet it is useful to better understand the neural network performances over known events. 
The splitting procedure divides the dataset in 75\% train, 25\% test; 30\% of the training set is further kept as validation set. The resulting splitting is: 52\% train, 23\% validation and 25\% test. The instances inside these sets are not sequential but rather taken randomly. \\

The purpose of our framework is to evaluate the effectiveness of our model on a known dataset, hence the choice of a loss function $L$ which is robust against outliers is critical. The Mean Square Error loss function (MSE) is:
\begin{equation}
\text{MSE}(x, y) = \frac{1}{n} \sum_{i=1}^{n}\left ( y_{i} - x_{i} \right )^2.
\end{equation}
MSE is very sensitive to the discrepancy between the prediction and the target value, thus it is a bad choice when outliers are present in the training dataset. 
We remark that the filtering of catalog events is not enough to guarantee the optimization of the background estimator when using MSE. Indeed, anomalous events, which are not present in the GBM catalog, may be over-fitted when minimizing MSE; these events are the actual targets of our search.

The Mean Absolute Error (MAE) loss function is less sensitive to residuals:
\begin{equation}
\text{MAE}(x, y) = \frac{1}{n} \sum_{i=1}^{n} \mid y_{i} - x_{i} \mid
\end{equation}
In respect to MSE, MAE will result in better neural network performances when anomalous events are included in the training dataset. 
In the settings of multi-output regression, the overall loss $\mathcal{L}$ is define as the MAE average of the NN outputs:
\begin{equation}
\mathcal{L} =  \frac{1}{m} \sum_{j=1}^{m}( \text{MAE}(F(X)_j, Y_j) )
\end{equation}

\subsection{Trigger algorithm}\label{sec_trigger}

An efficient change-point and anomaly detection algorithm called FOCuS-Poisson (Functional Online CUSUM) \cite{ward2022poisson} is employed to find anomalous transients in Fermi CSPEC data relative to the NN estimates of the background. \\
The FOCuS-Poisson algorithm is executed sequentially over the time series of the observed count data and the background estimates, separately for each combination of detectors and energy range. For a given detector-energy range combination with label $i$ and a given time step $t$, FOCuS-Poisson outputs an estimate of the maximum significance in the observed count excess relative to the background, $m_t^{(i)}$. This value is computed over an optimal time interval ending at $t$ and starting at a past time-step $t - d$. Crucially, the interval length $d$ is not predetermined but rather assessed and optimized by the algorithm itself, conditionally on the observations.
The significance values $m_t^{(i)}$ are recorded, in units of standard deviations, in a table with dimensions $M \times N$, where $M$ equals the length of the input time series and $N$ equals the number of detector-energy range combination. From these table, candidate transients are extrapolated in two steps. The first step is to identify vertical table slices (rows \textit{segments}) where a trigger condition is verified. The second step is to cluster together segments whose start and end times are closer than a pre-defined value (nominally $600$ s).\\
The user controls the search's output through three parameters. For the trigger condition to be verified it is required that the significance values exceed a threshold parameter $T$ over a minimum number detectors and energy ranges. Additionally, the user can limit the choice of the best interval to those whose length does not exceed a value $d_{\text{max}}$ or whose average intensity, given as a multiplicative factor of the observed counts in relation to the integral of background values, is greater than a minimum $\mu_{\text{min}}$. \\


\section{Data}\label{sec_data}

The Fermi/GBM daily \verb|CSPEC| data products were used for both the testing and the training of the neural network and for searching astrophysical transient events with FOCuS-Poisson. 
These data are photon counts with duration $4.096$~s, binned over $128$ logarithmically spaced energy channels spanning from  $\approx 8$ keV to $\approx 900$ keV \cite{meegan2009fermi}. 
The time resolution provided by \verb|CSPEC| data is high enough to investigate long and ultra-long GRBs, yet it is too low to reliably identify short GRBs and other transients with characteristic duration shorter than a few seconds. 
This is unfortunate, yet justified for our use-case. Indeed, the variability of background over time intervals of duration comparable to the duration of short GRBs is negligible, hence our method provides little benefits relative to simpler approaches such as moving average or exponential smoothing.
On the other hand, an accurate description of the background become essential when searching for long, faint events, in particular events whose duration is comparable to that of the Fermi orbit.
We consider data from all of the Fermi/GBM's twelve NaI detectors. Each detector is identified according to the standard GBM nomenclature (ten detectors labelled with integers ranging from $0$ to $9$, two detectors are identified by the letters $a$ and $b)$. 
In our analysis we disregard the Fermi/GBM bismuth-germanate detectors. These instruments are in fact sensible to energies much greater than the energies typically involved with GRBs prompt emission and are mainly used for the detection and observation of phenomena different from GRBs, such as Terrestrial Gamma-Ray Flashes (TGF) \cite{von2020fourth}.
To build the target variables $Y$, the input \verb|CSPEC| data from each detector are binned anew, this time over three coarser energy ranges (28-50 keV, 50-300 keV and 300-500 keV, see Table~\ref{tab:range}). The resulting dataset is arranged in a table with $36$ columns, one for each of the $36$ detector-energy combinations.

\begin{table}[!htb]
\centering
\begin{tabular}{l|r}
Range & Energy range (keV) \\\hline
r0 & 28-50 \\
r1 & 50-300 \\
r2 & 300-500
\end{tabular}
\caption{\label{tab:range} 
Energy ranges table. 
}
\end{table}

Beside the \verb|CSPEC| data product, the neural network is trained using information on the satellite geographical location and the detectors pointing direction, as well as a number of auxiliary features such as the Earth occultation status and the visibility of the Sun for each detector at a given time. These informations are gathered from the Fermi/GBM \verb|POSHIST| data products. A detail of the orbital and detectors features used in the training of the NN is given in Table~\ref{tab:feature} and Table~\ref{tab:feature_det}. 
To access both the \verb|CSPEC| and \verb|POSHIST| data products we use the Fermi Data Tools package, a python software API to the HEASARC Fermi archival database.
The resulting input datasets $X$ include a total of 60 different features, sampled with a step length of $4.096$ s.

\begin{table}[!htb]
\centering
\begin{tabular}{c|c}
Feature label & Description \\\hline
$pos\_x$, $pos\_y$, $pos\_z$ & position of Fermi in Earth inertial coordinates \\
$a$, $b$, $c$, $d$ & Fermi attitude quaternions \\
$lat$ & Fermi geographical latitude \\
$lon$ & Fermi geographical longitude \\
$alt$ & Fermi orbital altitude \\
$vx$, $vy$, $vz$ & velocity of Fermi in Earth inertial coordinates \\
$w1$, $w2$, $w3$ & Fermi angular velocity \\
$sun\_vis$ & Sun's visibility boolean flag.\\
$sun\_ra$ & Sun's right ascension\\
$sun\_dec$ & Sun's declination \\
$earth\_r$ & Earth's apparent radius \\
$earth\_ra$ & Earth center right ascension\\
$earth\_dec$ & Earth center declination \\
$saa$ & SAA transit boolean flag\\
$l$ & approximate McIlwain L value
\end{tabular}
\caption{\label{tab:feature} 
A table of the 24 orbital features used to form the NN's input table. The features are obtained from the POSHIST files and processed by the library Fermi GBM Data Tools.
}
\end{table}

\begin{table}[!htb]
\centering
\begin{tabular}{c|c}
Features & Description \\\hline
$ni\_ra$ & $i$-labelled detector pointing right ascension \\
$ni\_dec$ & $i$-labelled detector pointing declination \\
$ni\_vis$ & $i$-labelled detector Earth occulation boolean flag\\
\end{tabular}
\caption{\label{tab:feature_det} 
A table of the 36 detector features used to form the NN's input table. The features are obtained from the POSHIST files and processed by the library Fermi GBM Data Tools.
}
\end{table}


\section{Results}\label{sec_results}
In this section we present the results of the background estimator and the trigger algorithm application.
The open source code implementation is available on \href{https://github.com/rcrupi/DeepGRB}{github.com/rcrupi/DeepGRB}.

\subsection{Background estimator performance}\label{sec_results_nn}
\begin{figure}[!htb]
\centering
\includegraphics[width=1.\textwidth]{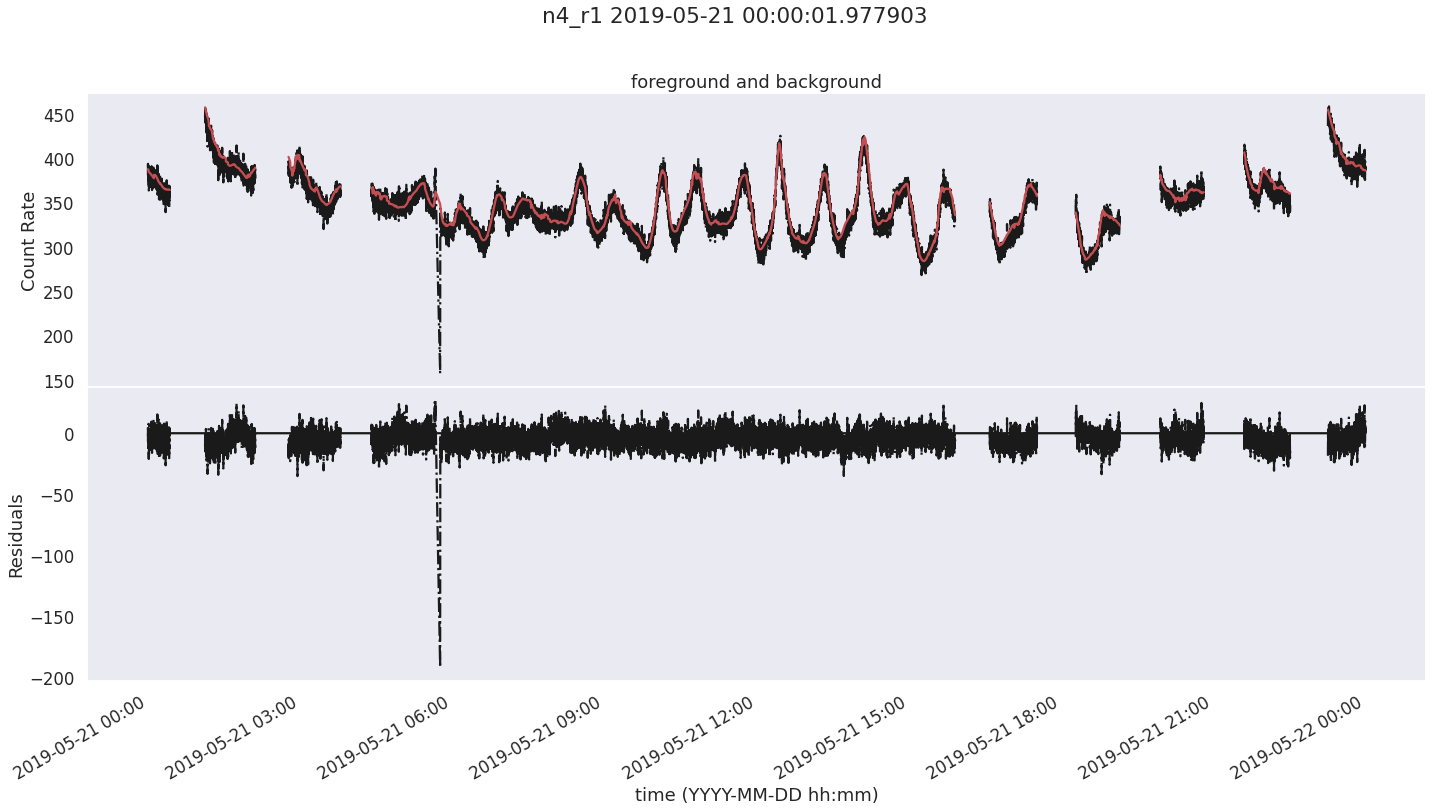}
\caption{\label{fig:residual}
The background estimation for the \texttt{n4} detector, in the energy range 1, on one day orbit. The Fermi/GBM count rate observations are represented over time as a black line, whereas the neural network estimation is plotted as a red solid line. The lower panel shows the residuals between the two quantities, with a black solid line denoting the reference of null residual.}
\end{figure}

To show the effectiveness of this approach, a NN is trained over 7 months of data from January to July 2019. An excerpt of the resulting background estimation is presented in Figure \ref{fig:residual} for one detector-range combination. The MAE values are reported in Table~\ref{tab:mae_summary}. The energy range bins are the same as those used in Section \ref{sec_data} and are defined in Table \ref{tab:range}.

\begin{table}[!htb]
\centering
 \begin{tabular}{||c | c c ||} 
 \hline
 det range & MAE train & MAE test \\
 \hline\hline
 r0 & 4.942 $\pm$ 0.331 & 4.953 $\pm$ 0.328  \\
 r1 & 6.088 $\pm$ 0.167 &  6.098 $\pm$ 0.163 \\ 
 r2 & 1.790 $\pm$ 0.044 & 1.792 $\pm$ 0.045 \\
 \hline
 average & 4.273 & 4.281 \\ 
 \hline
 \end{tabular}
 \caption{The NN MAE loss function (within one standard deviation) per energy range, over the training and the testing datasets, averaged over the 12 Fermi's GBM NaI detectors.
 }\label{tab:mae_summary}
\end{table}

In Appendix \ref{benchmark}, the NN background prediction over a dataset comprising GRB091024 are compared to the results of an established physical Fermi/GBM background model \cite{biltzinger2020physical}. 

\begin{figure}[!htb]
\hspace*{-1cm}   
\centering
\includegraphics[width=1.25\textwidth]{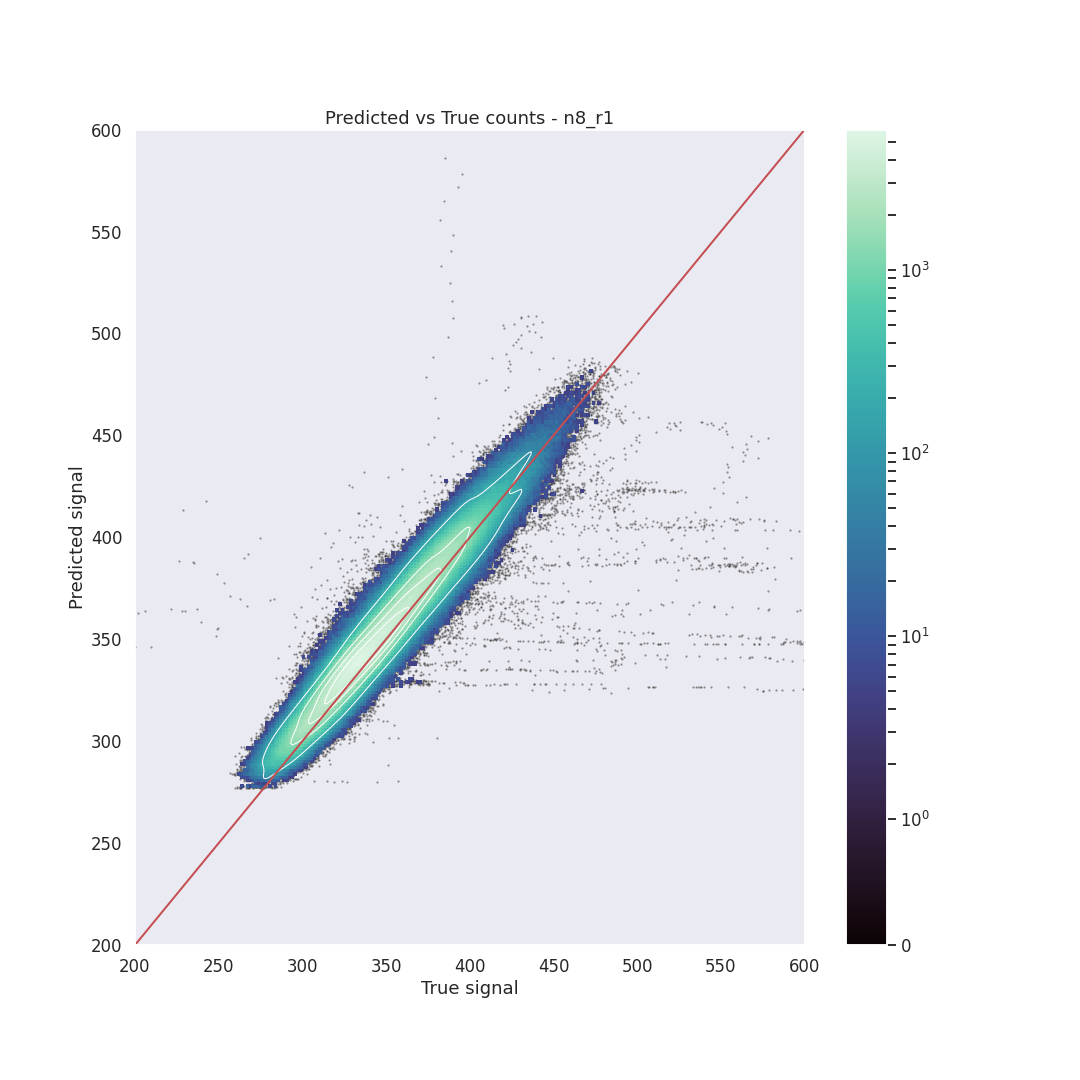}
\caption{\label{fig:bkg_est} Fermi/GBM photons counts from NaI-8 detector in the energy range $50$ - $300$ keV versus the respective prediction from the NN over the same combination of detector and energy range. Data spans from 1 January 2019 to 1 July 2019.}
\end{figure}

In Figure \ref{fig:bkg_est} we plot the NN predictions against the corresponding observed value, in particular we filtered out the data points 150 s before and after the SAA. 
Figure \ref{fig:bkg_est} depicts a light-curve with background counts greater than those observed, we discuss this phenomenon later in this Section \ref{sec_analysis}. 

\begin{figure}[!htb]
\centering
\includegraphics[width=.8\textwidth]{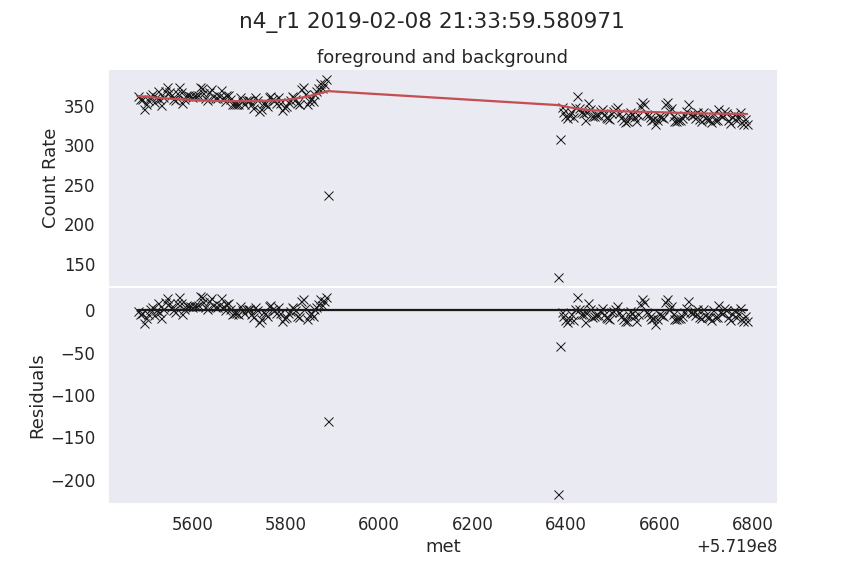}
\caption{\label{fig:switch}
Fermi/GBM NaI-4 detector photon count rates (crosses) in the energy range $50$ - $300$ keV versus the respective prediction from the Neural Network (red solid lines). The lower panel shows the residuals between the two quantities, with a black solid line denoting the reference of null residual. Data span $1400$ s and one SAA crossing. Three data events appear to be anomalously low, possibly due to lengthy instrument switch-on and switch-off procedures.}
\end{figure}

The events of interest for this research should be found when the observed count rates exceed the NN prediction, that are the points below the bisector (see Figure \ref{fig:bkg_est}) for detectors-range combination.
Transient, bright events such as GRBs may result in a temporary increase of the observed count rates and, taking place at random times and directions, are not predictable from features intrinsic to the Fermi spacecraft motion and attitude, which are the actual inputs of the NN. 

The three time periods chosen for the application of the trigger algorithm spans 1 November 2010 to 19 February 2011, 1 January 2014 to 28 February 2014, 1 March 2019 to 9 July 2019. For the sake of brevity, we will refer to these epochs as the `2010',  the `2014' and  the `2019' \emph{periods}.
These \emph{periods} are chosen to test the framework under a variety of conditions, including solar activity intensities and potential detector degradation.

A separate NN is trained and tested for each of these periods to account for variations in background count rates over long time scales (years), which may be caused by factors such as the two previously mentioned. We report the performance metrics in Table \ref{tab:nn_perf_table}.

\begin{table}[!htb]
\centering
\begin{tabular}{ |p{1cm}|p{1cm}||p{2.1cm}|p{2.1cm}| }
 \hline
 \multicolumn{4}{|c|}{NN performance metrics on test set} \\
 \hline
 Period & Energy range & MAE & MeAE \\
 \hline
 \hline
 2010 & r0 & 7.730 $\pm$ 4.842 & 3.963 $\pm$ 0.232 \\
 2010 & r1 & 6.469 $\pm$ 1.517 & 4.777 $\pm$ 0.100 \\
 2010 & r2 & 1.864 $\pm$ 0.033 & 1.563 $\pm$ 0.028 \\
 \hline
 2014 & r0 & 19.79 $\pm$ 18.92 & 4.545 $\pm$ 0.409 \\
 2014 & r1 & 13.29 $\pm$ 11.10 & 5.604 $\pm$ 0.196 \\
 2014 & r2 & 1.949 $\pm$ 0.099 & 1.598 $\pm$ 0.082 \\
 \hline
 2019 & r0 & 4.831 $\pm$ 0.300 & 3.938 $\pm$ 0.245 \\
 2019 & r1 & 5.640 $\pm$ 0.082 & 4.716 $\pm$ 0.070 \\
 2019 & r2 & 1.804 $\pm$ 0.038 & 1.510 $\pm$ 0.032 \\
 \hline
 \hline
\end{tabular}
 \caption{\label{tab:nn_perf_table} Mean Absolute Error (MAE) and Median Absolute Error (MeAE), on the test datasets, for each energy range and averaged for detectors within one standard deviation.}
\end{table}

Additional details on the neural network's performance during times of both high and low solar activity are provided in Appendix \ref{solarmaxmin}.


\subsection{Transients detection}\label{sec:trigstat}

With reference to the technique described in Section \ref{sec_trigger}, the following detection parameters were used to obtain the results discussed in this section.
The trigger condition was defined to resolve whenever at least one detector observed enough counts for the significance level to exceed a threshold $T = 3\sigma$ over the range of energy spanning $50$ keV and $300$ keV. 

To measure the consistency of the entire event it is computed the Standard Score $z$:
\begin{equation}
    z = \frac{x - \mu}{\sigma}
\end{equation}
where $\mu$ is the mean and $\sigma$ the standard deviation of the distribution $\mathcal{X}$.
Since we are dealing with count rates that follows the Poisson distribution, with sufficiently high count rates we can consider $\mu \approx \sigma^2$. Then the Standard Score can be approximated to:
\begin{equation}\label{eq:significance}
    S = \frac{N - B}{\sqrt{B}}
\end{equation}
where $N$ is the observed count rates integrated over an interval spanning the event's start time and end time\footnote{To avoid noise count rates and calculate the significance around the event's peak, only count rates greater than a quantile-based threshold were included in the integral.} and over each triggered detectors. Analogously for $B$ but the total count rates comes from the background estimated by the NN.
Standard Score is determined independently for each energy range $S_{r0}$, $S_{r1}$ and $S_{r2}$. The overall consistency for the event is defined as:
\begin{equation}
    C=max(S_{r0},\; S_{r1}, \; S_{r2}).
\end{equation}

\begin{figure}[!htb]
\centering
\includegraphics[width=1.\textwidth]{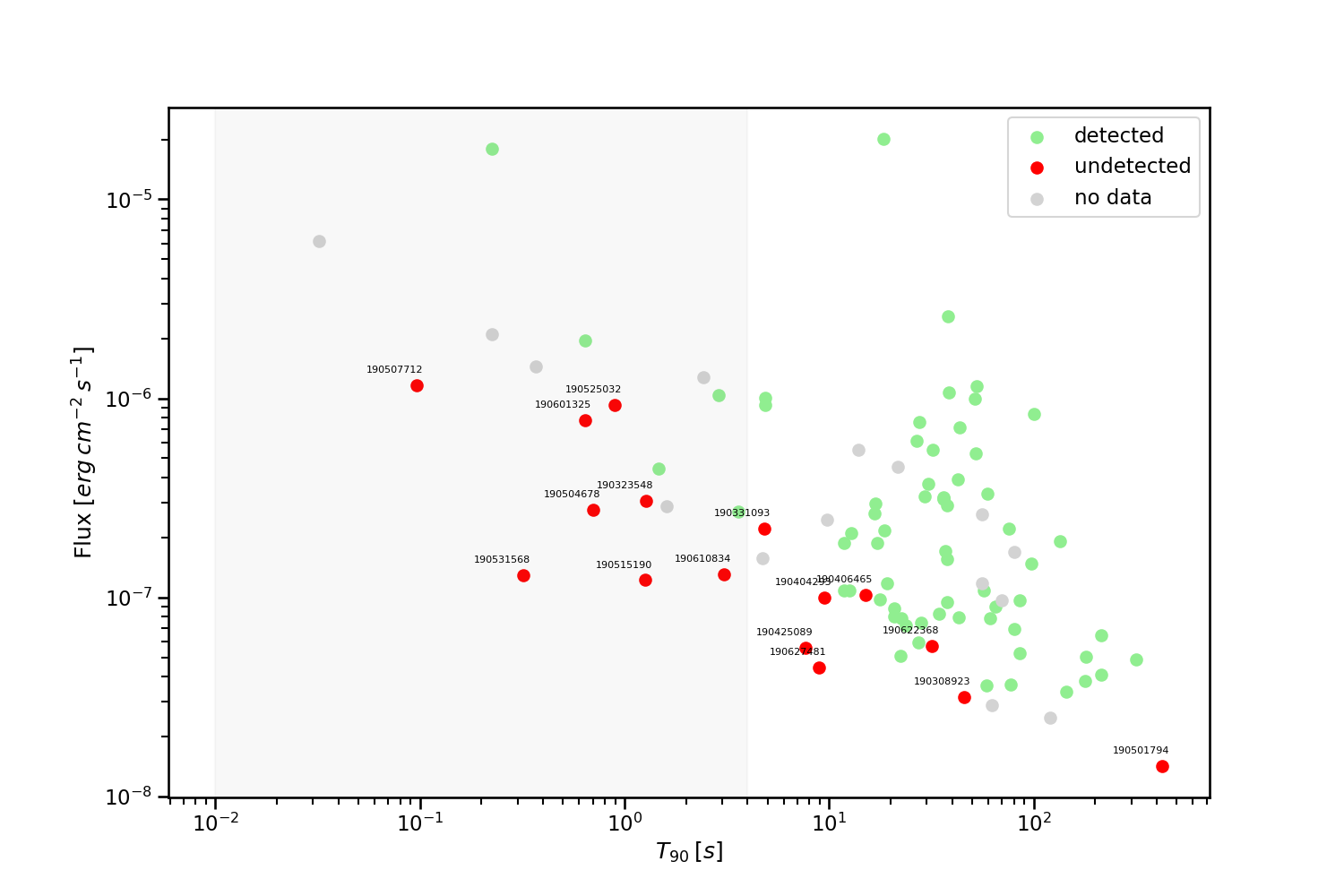}
\caption{\label{fig:red_green} GRB detection performances. Each dot represents a gamma-ray burst of the Fermi/GBM Burst Catalog discovered between March 1st and July 1st 2019 over the space spanned by the GRB's duration $T_{90}$ and flux, the latter computed as the ratio between the catalog's GRB fluence in band 10-1000 keV and $T_{90}$). Events in the shaded grey region have $T_{90}$ duration smaller than the bin-length time resolution of the dataset tested with the present framework ($4.096$ s, CSPEC data). Colors are used to identify the detection status within our search. In red the events unidentified with our method. Missing events (no data) are due to clipping of data 150 s before and after a SAA transit or portion of data that could not be preprocessed.}
\end{figure}

The FOCuS-Poisson algorithm was executed with the parameters $d_{\text{max}}$ and $\mu_{\text{min}}$  set to the values $120.4$ s and $1.2$ s, respectively. The choice of these parameters was driven by a trade-off between the need to find most astrophysical transients in our dataset both known and potentially unknown and the need to minimize the rate of false detection.
\\ 
The transient search was performed over three distinct time periods, as defined in the previous section.
In the period spanning March 2019 and July 2019 a total of 100 events were identified. Of these, 75 events match the trigger time of events already in the Fermi/GBM Trigger Catalog \cite{von2020fourth}, one event is due to artifacts in the dataset, while the nature of the remaining 24 events is uncertain. These results, along other from the remaining test periods, have been summarized in Table \ref{tab:detections_stats}.
Over the same period, the Fermi/GBM Burst Catalog \cite{von2020fourth} reports on 96 known GRBs. 
Of these bursts, 15 are missing a counterpart in our dataset due to the clipping of data $150$ s before and after a SAA transit. 
Of the remaining 76 bursts, 68 have $T_{90}$ duration larger than the bin-length resolution of our dataset ($4.096$ s). 
We were able to correctly identify $60$ of these bursts ($88 \%$).
Finally, we detected 5 out of 13 ($34\%$) GRBs with $T_{90}$ duration inferior to the the bin-lenght resolution of our dataset.
These results are summarized in Figure \ref{fig:red_green} and Table \ref{tab:burstcat_stats}, the latter also reporting on results from other periods.

\begin{table}[!htb]
\centering
\begin{tabular}{ |p{1cm}||p{1.2cm}|p{1.2cm}|p{1.5cm}|p{1.2cm}|p{1.5cm}|p{1.2cm}| }
 \hline
 \multicolumn{7}{|c|}{Transient detection statistics} \\
 \hline
 Period & Total events & Known & Consistency median known & Unknown & Consistency median unknown  & False Detections \\
 \hline
 2010 & 81   & 58 & $>10$ & 15 & 8.83 & 8 \\
 2014 & 195  & 81  & $>10$  & 71 & $>10$  & 43\\
 2019 & 100  & 75  & $>10$ & 24  & 8.71  & 1\\
 \hline
\end{tabular}
 \caption{\label{tab:detections_stats}  Total number of transients identified, number of transients with counterparts in the Fermi/GBM Trigger Catalog and its median consistency; number of transients of uncertain origin with no counterparts in the Fermi/GBM Trigger Catalog and its median consistency; false detection. Each table row corresponds to a different time period.}
\end{table}

\begin{table}[!htb]
\centering
\begin{tabular}{ |p{1cm}||p{1.5cm}|p{1.5cm}|p{2cm}|p{2cm}| }
 \hline
 \multicolumn{5}{|c|}{Known GRBs detection statistic} \\
 \hline
 Period & Total GRBs & Missing (no data) & $T_{90} > 4.096$ s & $T_{90} < 4.096$ s\\
 \hline
 2010 & 77  & 11  & 39/52 (75 \%)  & 2/14 (14 \%)\\
 2014 & 36   & 8 & 17/18 (94 \%) & 4/10 (40 \%) \\
 2019 & 96  & 15 & 60/68 (88 \%) & 5/13 (34 \%) \\
 \hline
\end{tabular}
 \caption{\label{tab:burstcat_stats} Total number of Fermi/GRBs in the Fermi/GBM Burst Catalog, number of events with missing (no data) counterparts in our dataset due to SAA data clipping (see Section \ref{sec:trigstat}), fraction of detected bursts with duration greater than the bin-length time resolution of the tested dataset, fraction of detected bursts with duration smaller than the bin-length time resolution of the tested dataset. Each table row corresponds to a different time period.}
\end{table}

\section{Discussion}\label{sec_analysis}

\begin{figure}
\centering
\begin{subfigure}{0.45\textwidth}
\centering
\includegraphics[width=\linewidth]{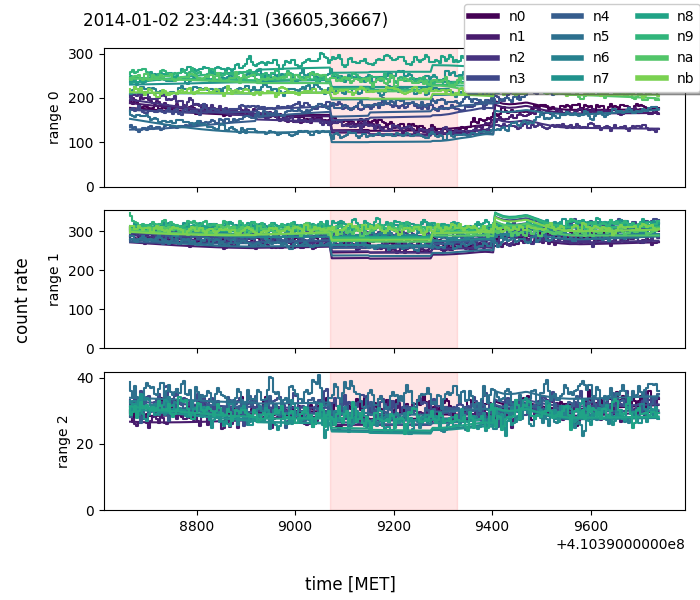}
\caption{\label{fig:erronn} 
}
\end{subfigure}
 \hspace*{\fill} 
\begin{subfigure}{0.45\textwidth}
\centering
\includegraphics[width=\linewidth]{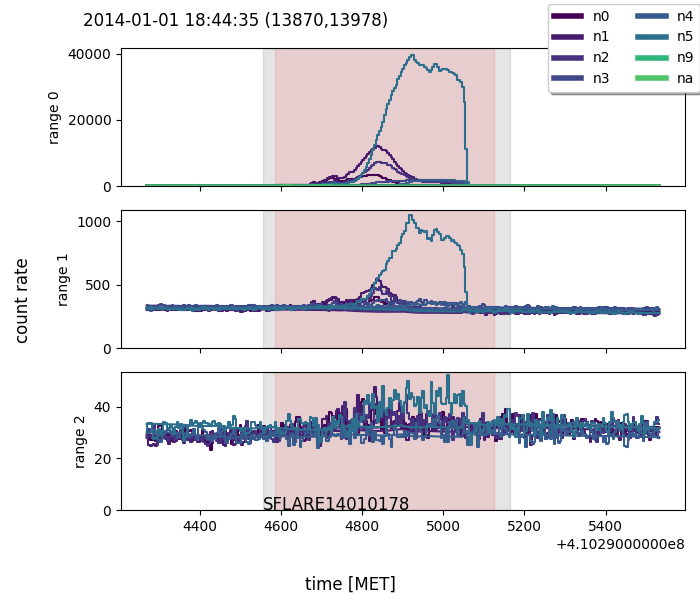}
\caption{\label{fig:bigsolarflare} 
}
\end{subfigure}

\caption{Photon counts from each triggered detector are plotted with step lines, across three energy bands spanning $28-50$ keV, $50-300$ keV and $300-500$ keV (Table \ref{tab:range}), with a resolution of $4.096$~s. 
The neural network's prediction of background count rates is represented by solid lines.
Different detectors are identified using different colors.
 A red shaded area limits FOCuS-Poisson's best guess of the transient duration. Times are expressed in units of seconds according to Fermi's standard mission elapsed time (MET).
(a) Example of False Detection in which all the detector are triggered over an imprecision of the Neural Network estimation. (b) Example of a solar flare in the Fermi/GBM catalog detected by our approach. The event start and end MET time, as reported in the Fermi/GBM trigger catalog, is represented by a grey shaded area.} \label{fig:2014errsf}
\end{figure}

According to Tables~\ref{tab:mae_summary} the test set and train set MAE values are similar up to $1 \%$ indicating no over-fitting and strong generalization across energy range and detector.
Table \ref{tab:nn_perf_table} shows that the neural network trained on data from the 2014 period has the highest (worst) MAE, which can be attributed to the presence of strong solar activity. This is understandable since, during an activity maximum, the background particle count rate is more unpredictable due to the influence of the Sun on the local radiation environment (see Figure \ref{fig:bigsolarflare}). Nonetheless, MeAE shows similar performance with the other two periods, thanks to its robustness against outliers. On the other hand, the 2019 period has the lowest MAE most likely due to low solar activity and low background variability. For more details on predictive robustness against high solar activity see Appendix \ref{solarmaxmin}. 

In Figure \ref{fig:bkg_est}, most of the data points are distributed along the plot bisector $y = x$, indicating that most often the neural network estimate is in agreement with the actual observations.
Above the bisector, more counts are expected than they are actually observed. From spot analysis of these datapoints it is apparent that overestimates in NN prediction often happen when Fermi enters or leaves the SAA, see for example Figure \ref{fig:switch}. 
A possible explanation of this behaviour can be rooted in the Fermi/GBM experimental apparatus. Fermi/GBM is composed of multiple photomultipliers. These instruments require high voltages to operate. A ``ramp-up" of a photomultiplier to operational voltage takes place over a time span of several seconds. During this time, the photomultiplier amplification factor (the gain) is hindered resulting in lower than nominal count rates. The same effect takes place when ``ramping-down" before turning a photomultiplier off. In the lower part of the bisector the observed counts rate exceed the counts estimated. The trigger algorithm's goal is to determine whether this exceeding amount is part of an event or simply a random fluctuation.

In periods of high solar activity, Fermi/GBM data include a large number of soft transient events of solar origin; thus, the soft ($25$ - $50$ keV) trigger conditions have been disabled on multiple occasions (e.g. see Table 4 in \cite{bhat2016third} for 2014). Likewise, we required that at least one detector must be over threshold in the energy band spanning $50$ and $300$ keV in order for the trigger condition to be satisfied. 
Still, Table \ref{tab:detections_stats} shows a higher number of total events for the 2014 period. The majority of these events are most likely associated to solar flares; indeed, 50 of the 81 events in the GBM trigger catalog for this period are solar flares, and the majority of the events we find with no counterpart in the Fermi/GBM trigger catalog are triggered over Sun facing detectors (n0, n1, n2, n3, n4, n5).
False detections may be caused by artifacts in the background estimation. These are generally easy to identify; most of the time these artifacts take the form of sudden steps in the background estimate, simultaneously over all detector/range combination. One of these events is represented in Figure \ref{fig:erronn}. This behaviour is less frequently present in the other two periods analyzed, indicating that noisy background impacts on performance (see MAE) and therefore more false positive are detected.

To further investigate the detected and undetected GRBs, we plot the flux (total fluence divided by T90) vs T90 for our triggered events in Figure \ref{fig:red_green}. The red points are events reported in the Fermi/GBM catalog but undetected by our method. The Fermi/GBM events with a duration less than our time binning (4.096s) are often undetected in our analysis because of the too coarse binning. We also miss a few longer events with low count rates. Reducing the time binning by using data with higher time resolution, such as \verb|CTIME| or \verb|TTE|, could be beneficial to capture shorter and fainter events. Despite the unfavorable adopted time binning of 4.096s, we recovered $\ge 75\%$ of the GRBs with $T_{90}$ greater than 4.096s, see Table \ref{tab:burstcat_stats}. 

We also detect many events not present in the Fermi/GBM catalog, 
and we use the methodology outlined in \cite{kommers1999faint} to characterize these transients. More specifically, we classify events as:
\begin{itemize}
\item Solar flare (SF) when the majority of the counts are in the low-energy range and the Sun is in the field of view of the triggered detectors. 
\item Terrestrial Gamma-ray Flash (TGF) when most of the counts are in the high-energy range and the event's  source reaches the detector from the Earth’s horizon.
\item Gamma-ray burst (GRB) when most of the counts are in the $50-300$keV energy range, and the source direction is not occulted by the Earth and is distant from both the Sun and the galactic plane.
\item Galactic X-ray flash (GF) when the source direction is compatible with that of the galactic plane.
\item Uncertain (UNC) in all other cases. 
\end{itemize}
To determine the source direction, we employ a simple method based on the evaluation of the pointing and the relative photon count rate of the detectors. Further details can be found in Appendix \ref{localization}. 

Two classes of transient events are discussed further in this section: events already classified as GRBs in the Fermi/GBM trigger catalog; events not present in the Fermi/GBM catalog but classified by us as candidate GRBs. We report In Table \ref{tab:events} six more events that have no catalog counterpart, suggesting one or more of the previously mentioned categories. All these events are a cherry pick selection of the unknown events in Table \ref{tab:catalog_unk}.

\subsubsection*{GRB 190320A}
\begin{figure}[!htb]
\centering
\includegraphics[width=0.75\textwidth]{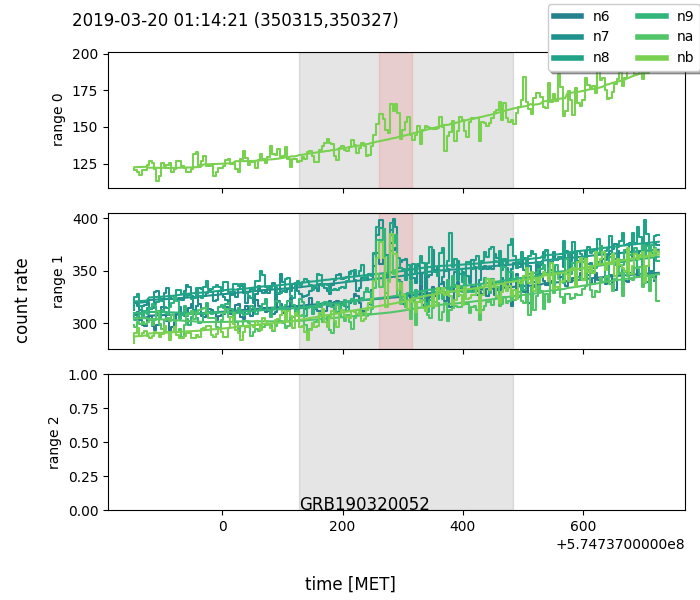}
\caption{\label{fig:grb20190320} The Fermi/GBM catalog GRB190320, as detected by our method.
Photon counts from each triggered detector are plotted with step lines, across three energy bands spanning $28-50$ keV, $50-300$ keV and $300-500$ keV (Table \ref{tab:range}), with a resolution of $4.096$~s. 
The neural network's prediction of background count rates is represented by solid lines.
Different detectors are identified using different colors.
The GRB start and end MET time, as reported in the Fermi/GBM burst catalog, is represented by a grey shaded area. A red shaded area limits FOCuS-Poisson's best guess of the transient duration. Times are expressed in units of seconds according Fermi's standard mission elapsed time (MET).
}
\end{figure}

At 01:14:16 UTC on March 20, 2019, the long GRB 190320052 triggered the Fermi/GBM on board trigger algorithm across detectors \texttt{n6} and \texttt{n9}. The estimated $T_{90}$ duration is $43$~s, with the highest emission component in the $50$-$300$ keV band.
In our analysis, the detectors \texttt{n6}, \texttt{n7}, \texttt{n8}, \texttt{n9} and \texttt{na} all exceeded a $3.0$~$\sigma$ significance threshold during the period event (Figure \ref{fig:grb20190320})  with a resulting consistency greater than $10$ on energy range r1 and $5.74$ on r2. The background estimate is comparable to a second order polynomial fitting in the soft energy range and first order polynomial fitting in the $50$-$300$ keV energy range.

\subsubsection*{Event 190420939}
\begin{figure}[!htb]
\centering
\includegraphics[width=0.75\textwidth]{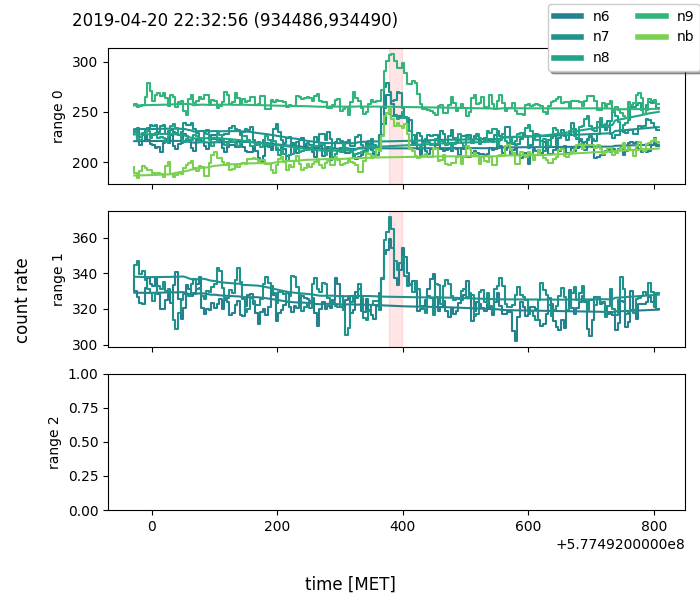}
\caption{\label{fig:grb20190420} The 190420939 transient event with no direct counter part in the Fermi/GBM trigger catalog. The event was classified as a candidate gamma-ray burst, according to the discussion presented in Section \ref{sec_analysis}. For the corresponding localization see Figure \ref{fig:loc20190420}.
Photon counts from each triggered detector are plotted with step lines, across three energy bands spanning $28-50$ keV, $50-300$ keV and $300-500$ keV (Table \ref{tab:range}), with a resolution of $4.096$~s. 
The neural network's prediction of background count rates is represented by solid lines.
Different detectors are identified using different colors. A red shaded area limits FOCuS-Poisson's best guess of the transient duration. Times are expressed in units of seconds according Fermi's standard mission elapsed time (MET).}
\end{figure}

Figure \ref{fig:grb20190420} shows an event not present in the GBM trigger catalog, similar to GRB190320052 but with higher low-energy count rate. 
The event has been triggered by detectors \texttt{n6}, \texttt{n7}, \texttt{n8}, \texttt{na} and \texttt{nb} in the low energy band with a consistency greater than $10$. Two detectors provided a trigger in the $50$-$300$ keV energy band, with a consistency of $8.4$.

\begin{figure}[!htb]
\centering
\includegraphics[width=1.0\textwidth]{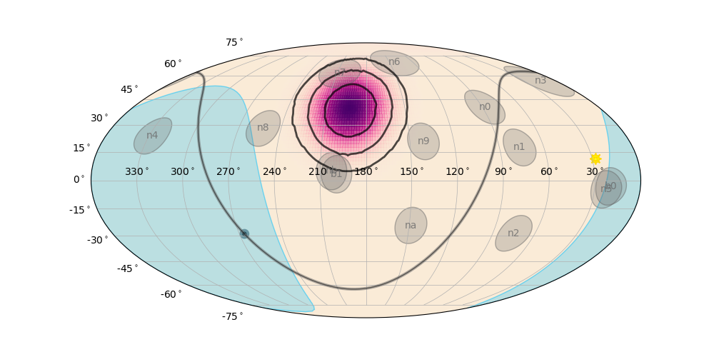}
\caption{\label{fig:loc20190420} Estimate of the candidate event's source localization over the celestial sphere at 2019-04-20 22:32:56 UTC.}
\end{figure}

We can see from the localization estimate in Figure \ref{fig:loc20190420} that the event is far from the galactic plane, the Sun, and the Earth's horizon. With all of this information, this event could be a long soft GRB. 
The localization algorithm used is described in detail in Section \ref{localization}.

\subsubsection*{Interesting events}

We list in Table~\ref{tab:events} a selection of interesting events, including the one already discussed, which are not present in the GBM catalog and which deserve further analysis. Appendix \ref{events} present plots associated to these events.
Events 1 and 2 are classified as Solar Flares because their location is close to the Sun and the majority of the detectors triggered are in the energy range r0.
Because event 3 is far from the Sun yet close to the galactic plane and the Earth's horizon, it might be a Galactic X-ray flash or a Terrestrial Gamma-ray Flash.
Event 4 and 6 are categorized as GRBs for the same reasons as event 5, however because they are near the galactic plane, event 6 might be a Galactic X-ray burst.
Finally, in event 7, nine detectors with roughly equal intensities are triggered, suggesting that this event is likely due to Local Particles. This is further validated by the satellite's position at high geomagnetic latitude (Figure \ref{fig:events7}), which is highly correlated with the localization of charged particle events \cite{von2020fourth}; as a result, the event is classified as uncertain.

It's worth noting that GRB 190404B \href{https://www.mpe.mpg.de/~jcg/grb190404B.html}{GCN Circular notice}
discovered by Monitor of all-sky X-ray image (MAXI) satellite\footnote{\href{http://maxi.riken.jp/grbs/190404b/}{http://maxi.riken.jp/grbs/190404b/}} has location $(\text{RA} = 221^{\circ}, \text{Dec} = -22^{\circ})$, which is similar to event 4, and trigger time 2019/04/04 13:14:34.00 UTC, which is six minutes after event 4.

The complete catalog of unknown and known events for the three time periods analyzed can be found in Appendix in Tables \ref{tab:catalog_unk} and \ref{tab:catalog_kn}, respectively. The events are reported with the trigger time, duration, the triggered detectors, the Standard Score for each energy range, and a significance classification. Unknown events were assigned tentative transient classes using the methodology described in this section.

\begin{table}[!htb]
\centering 
\begin{tabular}{c c c p{0.1\textwidth} c c p{0.1\textwidth} c}
 \multicolumn{8}{c}{} \\
ID & Trigger time & T (s) & Detectors triggered & RA ($^\circ$) & Dec ($^\circ$) & Transient class & C \\\hline
1 & 2014-01-27 05:21:12 &	32.77 &	n0 n1 n2 n3 n4 n5 n8 nb & 306 & -22 & SF & $>10$ \\
2 & 2010-11-11 18:58:17 &   16.38 &	n2 n4 n5 & 230 & -20 & SF & $>10$ \\
3 & 2014-01-12 13:59:58 &	102.40 &	n6 n7 n8 n9 na nb & 105 & 10 & GF/TGF &  $>10$ \\
4 & 2019-04-04 13:08:07 &   8.19 &	n9 na & 220 & -10 & GRB & 4.93 \\
5 & 2019-04-20 22:32:56 &  16.38 &	n6 n7 n8 n9 nb & 245 & 40 & GRB & $>10$ \\
6 & 2019-06-06 13:21:42 &   16.38 &	n7 n8 nb & 250 & 25 & GRB/GF & 9.12 \\
7 & 2011-02-15 15:59:02 &  118.79 & n0 n1 n2 n5 n6 n7 n8 n9 nb & 208 & 62 & UNC & $>10$

\end{tabular}
\caption{\label{tab:events} List of interesting events. We report the ID, the start time in MET and UTC, the end time in MET, the detectors triggered during the event, the localisation expressed in right ascension and declination, the proposed transient class  and the consistency of the event}.
\end{table}

\newpage

\section{Conclusion}\label{sec:conclusion}
A novel method for high-energy, transient event detection is presented, integrating the precise estimation of a NN with an efficient trigger algorithm. The method has been designed to be applied to HERMES Pathfinder data, but it can be extended to analyze data from other space-based, high-energy missions and we have presented here an application using Fermi/GBM data. The first step is to estimate the background count rate with a NN using satellite data that may be used to build a physical background model. The accuracy of the background estimate is measured using Mean Absolute Error and Median Absolute Error. An experiment is carried out to assess the robustness of the background estimator during the periods of solar maxima (2014) and solar minima (2020), demonstrating that the background estimation is stable enough to have comparable performance in both periods. Because HERMES Pathfinder will be deployed near the next solar maximum, a scenario of expected count rates is provided in Appendix.  The background is then used by FOCuS-Poisson, an evolution of the CUSUM algorithm, to efficiently detect the transient events.  This method is tested using three periods of Fermi/GBM data binned in time for 4s. We provide statistics on known and unknown transients in the GBM catalog. We show that with our method we are able to recover known events longer than 4s, and to selected events not included in the Fermi/GBM catalog. Seven of the unknown events are discussed in details.
Future work will aim to improve neural network prediction, such as employing a Recurrent Neural Network to provide a smoother signal or reducing time binning to detect shorter events. Also, provide explainability methods to allow the user to explain a specific prediction or debug the NN.

\section*{Acknowledgement}
Special thanks to Daniele Regoli for the useful recommendations in the introduction and method sections.
This research acknowledge support from the European
Union Horizon 2018 and 2020 Research and Innovation Frame-work Programme under grant agreements HERMES-Scientific Pathfinder n. 821896 and AHEAD2020 n. 871158, and by ASI INAF Accordo Attuativo n. 2018-10-HH.1.2020 HERMES—Technologic Pathfinder Attivita’ scientifiche. We also aknownledge the support of the INAF RSN-5 mini-grant 1.05.12.04.05, "Development of novel algorithms for detecting high-energy transient events in astronomical time series".

\bibliographystyle{unsrtnat}
\bibliography{references}

\newpage
\appendix

\section{Background estimate for GRB 091024}\label{benchmark}
To demonstrate the potential of the background estimator in the presence of a long event, a background estimation is performed in a period containing the ultra-long GRB 091024 \cite{gruber2011fermi}, for which a similar evaluation is provided in \cite{biltzinger2020physical}. 

\begin{figure}[H]
\centering
\begin{subfigure}{.3\linewidth}
    \centering
    \includegraphics[width=1.25\textwidth]{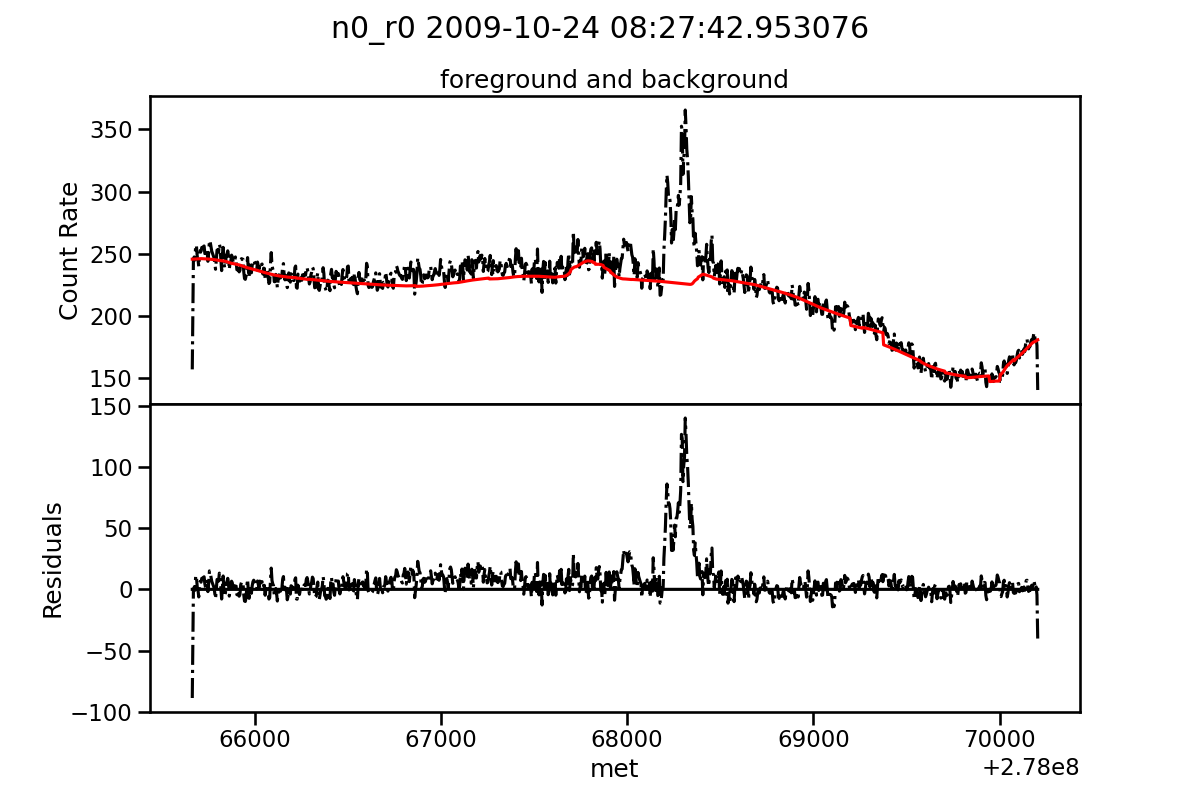}
\end{subfigure}
    \hfill
\begin{subfigure}{.3\linewidth}
    \centering
    \includegraphics[width=1.25\textwidth]{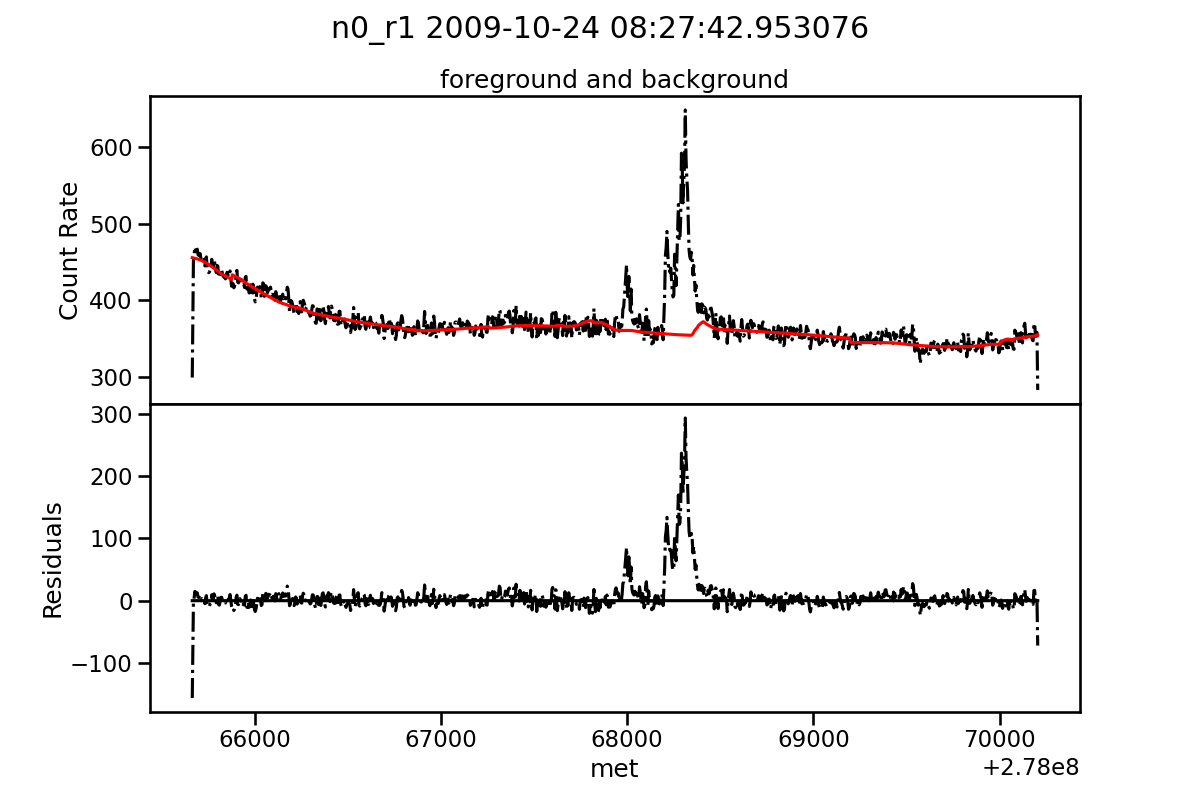}
\end{subfigure}
   \hfill
\begin{subfigure}{.3\linewidth}
    \centering
    \includegraphics[width=1.25\textwidth]{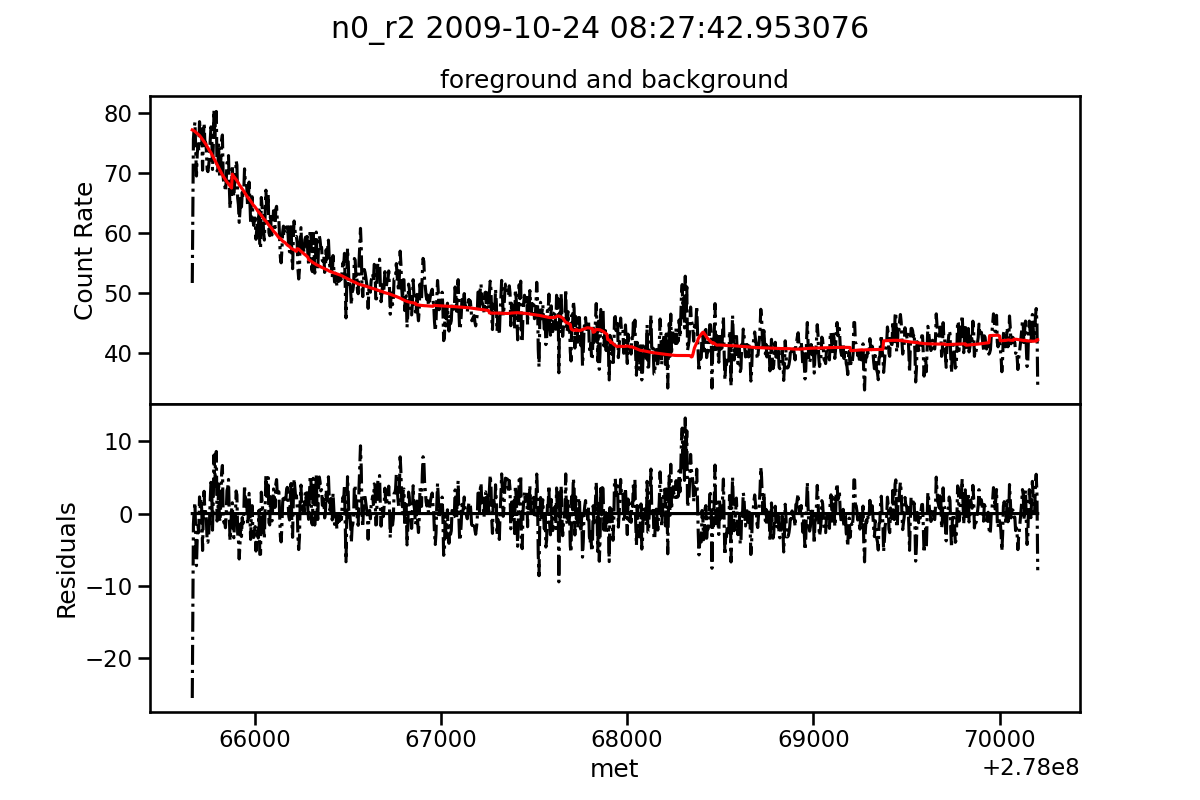}
\end{subfigure}

\bigskip
\begin{subfigure}{.3\linewidth}
    \centering
    \includegraphics[width=1.25\textwidth]{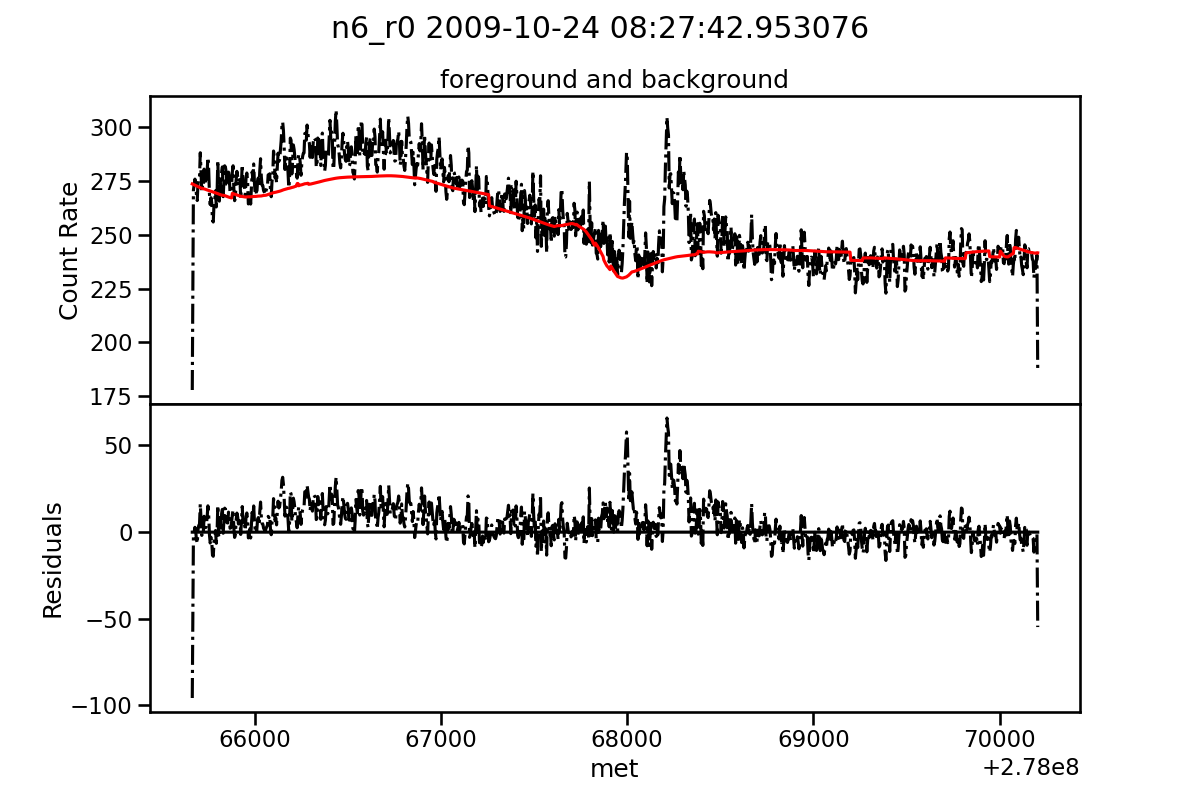}
\end{subfigure}
    \hfill
\begin{subfigure}{.3\linewidth}
    \centering
    \includegraphics[width=1.25\textwidth]{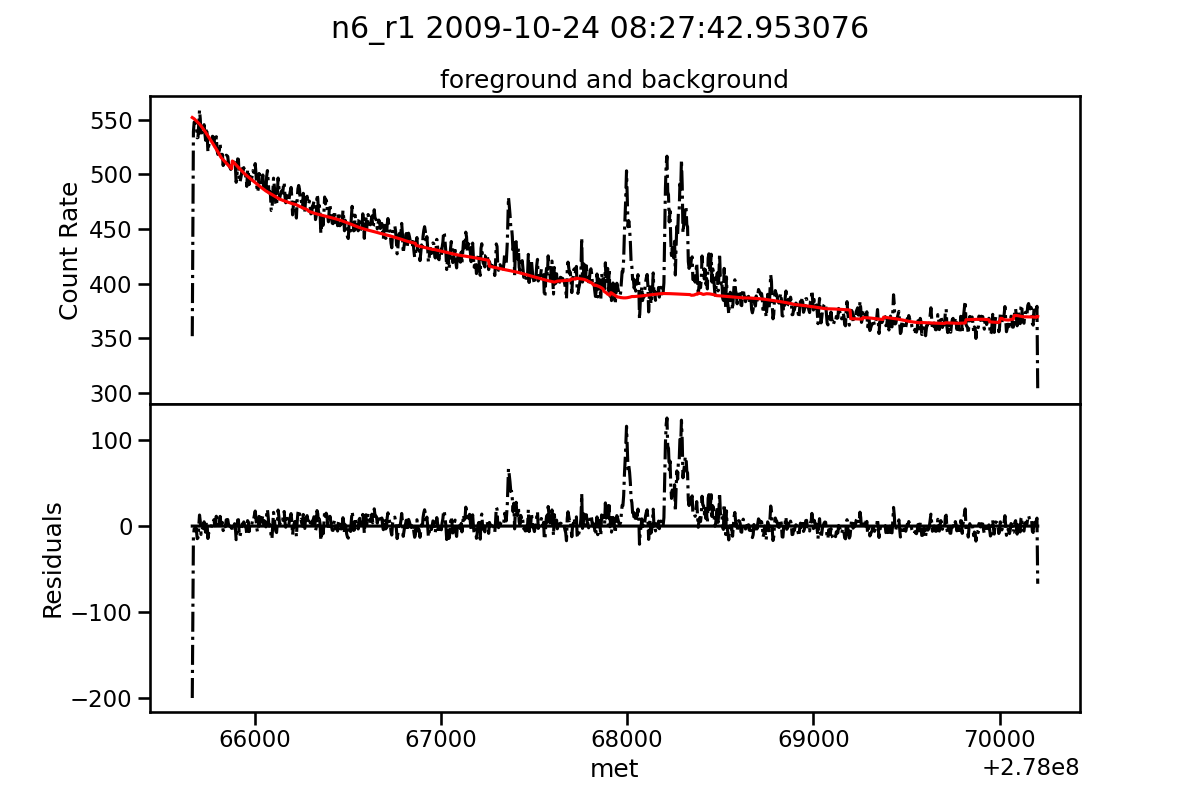}
\end{subfigure}
   \hfill
\begin{subfigure}{.3\linewidth}
    \centering
    \includegraphics[width=1.25\textwidth]{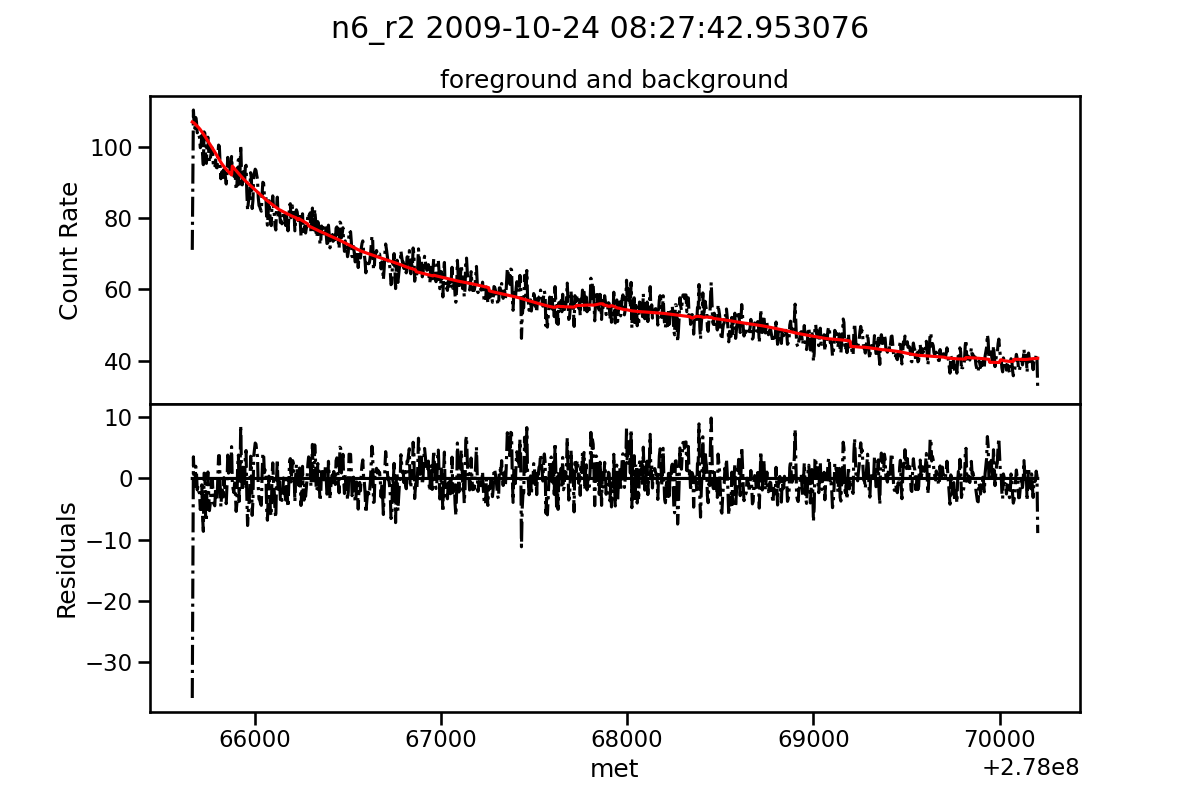}
\end{subfigure}

\bigskip
\begin{subfigure}{.3\linewidth}
    \centering
    \includegraphics[width=1.25\textwidth]{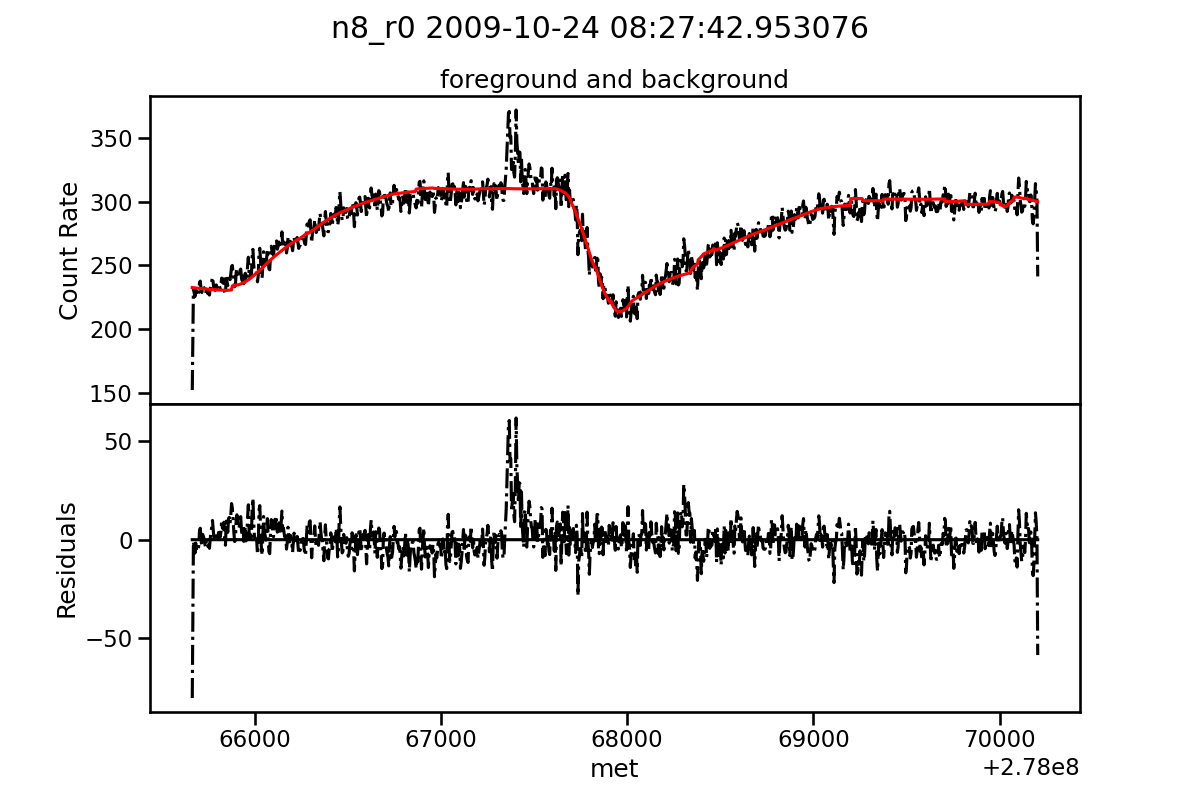}
\end{subfigure}
    \hfill
\begin{subfigure}{.3\linewidth}
    \centering
    \includegraphics[width=1.25\textwidth]{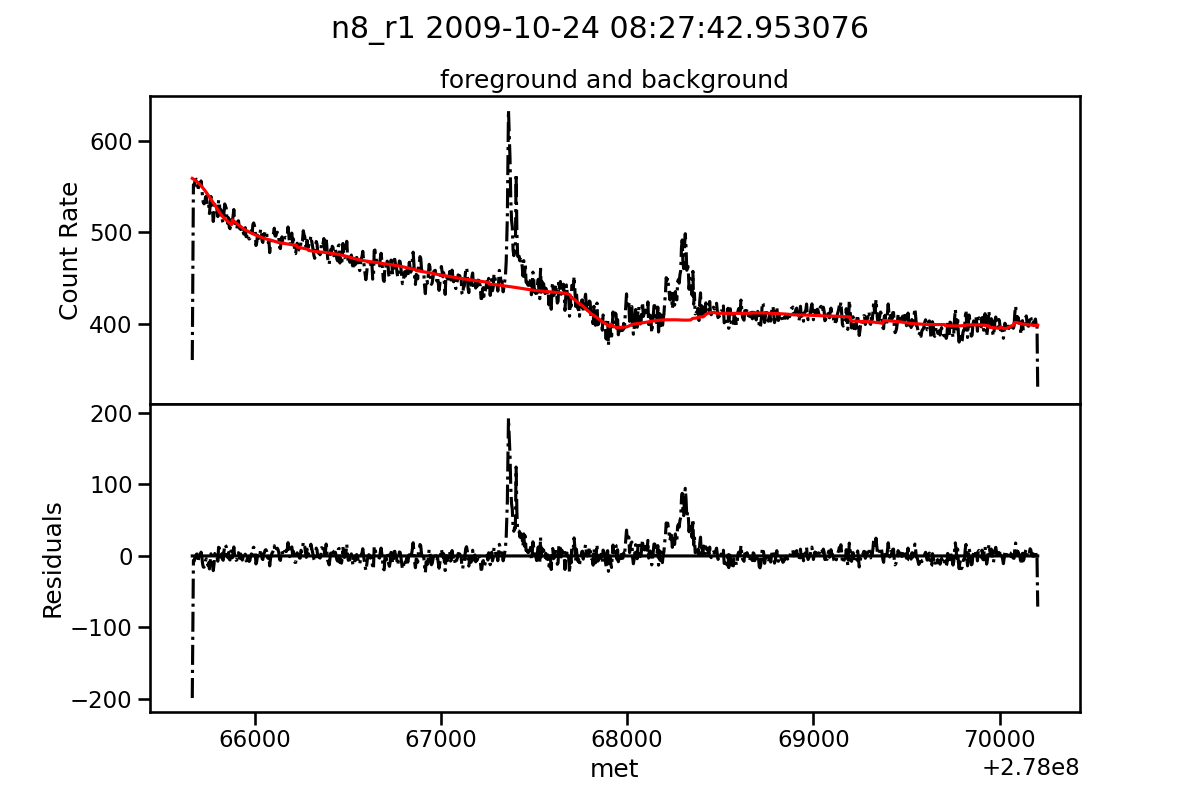}
\end{subfigure}
   \hfill
\begin{subfigure}{.3\linewidth}
    \centering
    \includegraphics[width=1.25\textwidth]{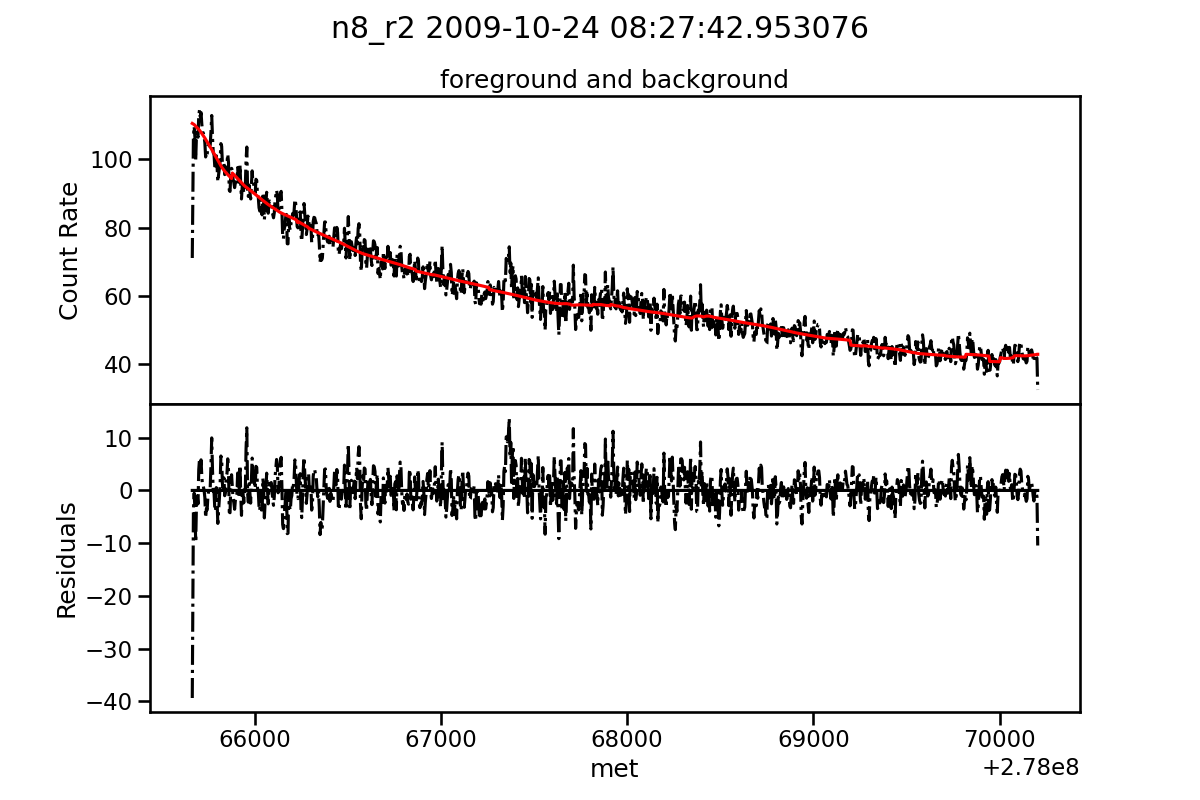}
\end{subfigure}

\caption{Observed and background estimate count rates around the event GRB 091024. From left to right the plots refer respectively to range r0, r1, r2, from top to bottom the plots refer respectively to detectors \texttt{n0}, \texttt{n6}, \texttt{n8}. This figure can be compared with Figure 18 in \cite{biltzinger2020physical}.
\label{fig:091024}}
\end{figure}

In Figure \ref{fig:091024} are shown detectors \texttt{n0}, \texttt{n6} and \texttt{n8} in the three energy band specified in Table~\ref{tab:range}.
The data and background estimation of a Neural Network trained and tested during a three-month period, from September 1 to November 30, 2009, are presented in black and red, respectively.
The dataset consists of 1.63 million of samples and the hyperparameters are the same used in Section \ref{sec_background} except for the learning rate
\begin{equation*}
  \eta =
    \begin{cases}
      0.025 & \text{if epoch $<4$}\\
      0.004 & \text{if $4\ge$ epoch $<12$}\\
      0.001 & \text{if epoch $\ge 12$}
    \end{cases}.       
\end{equation*}
The event emerges clearly from the residuals of all the detectors in range r1 and r2, in detector \texttt{n0} and range r2 it is still visible a peak probably belonging to the end of it. In detector \texttt{n6} range r0, in the first part of the time series before the peaks of the event, the background estimation underestimates the foreground (data observed). This could be due to a too short period of training dataset, a non optimal parameter settings of the NN, a different event such as Local Particles or, more interestingly, the first part of the GRB, where photon counts were too low to be detected due to background variability.

\section{Localization}\label{localization}
For the standard reference for the localization of events found by GBM, look \cite{goldstein2020evaluation}.\\
In this work the localization is done by a simple geometric reasoning, but in future we hope to use more sophisticated algorithm of localization. To optimise the function loss it is employed a particle swarm optimiser\footnote{\href{https://github.com/tisimst/pyswarm}{https://github.com/tisimst/pyswarm}}.

Consider two vectors in the equatorial coordinates $\psi_d = (ra_d, dec_d)$ and $\psi_s = (ra_s, dec_s)$, respectively the pointing of a detector and the localization of the event source. The incidence intensity is modeled as the cosine between the angle $cos(\psi_d, \psi_s)$ is:
\begin{align*}
    cos(\psi_d, \psi_s) & = cos(\psi_{d,ra}) cos(\psi_{d,dec}) cos(\psi_{s,ra}) cos(\psi_{s,dec}) + \\
    & + sin(\psi_{d,ra}) cos(\psi_{d,dec}) sin(\psi_{s,ra}) cos(\psi_{s,dec}) +  \\
    & + sin(\psi_{d,dec}) sin(\psi_{s,dec}) 
\end{align*}
              
If the angle of incidence is grater than $\pi/2$ than the incident intensity must be set to 0. Finally we have \ref{ii}
\begin{equation}\label{ii}
\mathcal{I} (\psi_d, \psi_s) = max(cos(\psi_d, \psi_s), 0)
\end{equation}

The loss to optimise in Equation \ref{loss_pos}, where $i$ is a particular detector in $D$ detectors (in our case 12).
The energy range chosen is the one with the biggest residuals among detectors/energy ranges, then the count rates corresponding to the timestamp of the maximum value is given to the loss \ref{loss_pos} and minimized.

\begin{equation}\label{loss_pos}
    \frac{\sum_{i=1}^{D} ( counts_s (\mathcal{I} (\psi_i, \psi_s) ) - counts_i )^2}{D}
\end{equation}
where $\psi_s$ and $counts_s$ are the unknown variables.

\section{Solar minima and maxima}\label{solarmaxmin}
\begin{figure}[H]
\hspace*{-2cm} 
\centering
\includegraphics[width=1.25\textwidth]{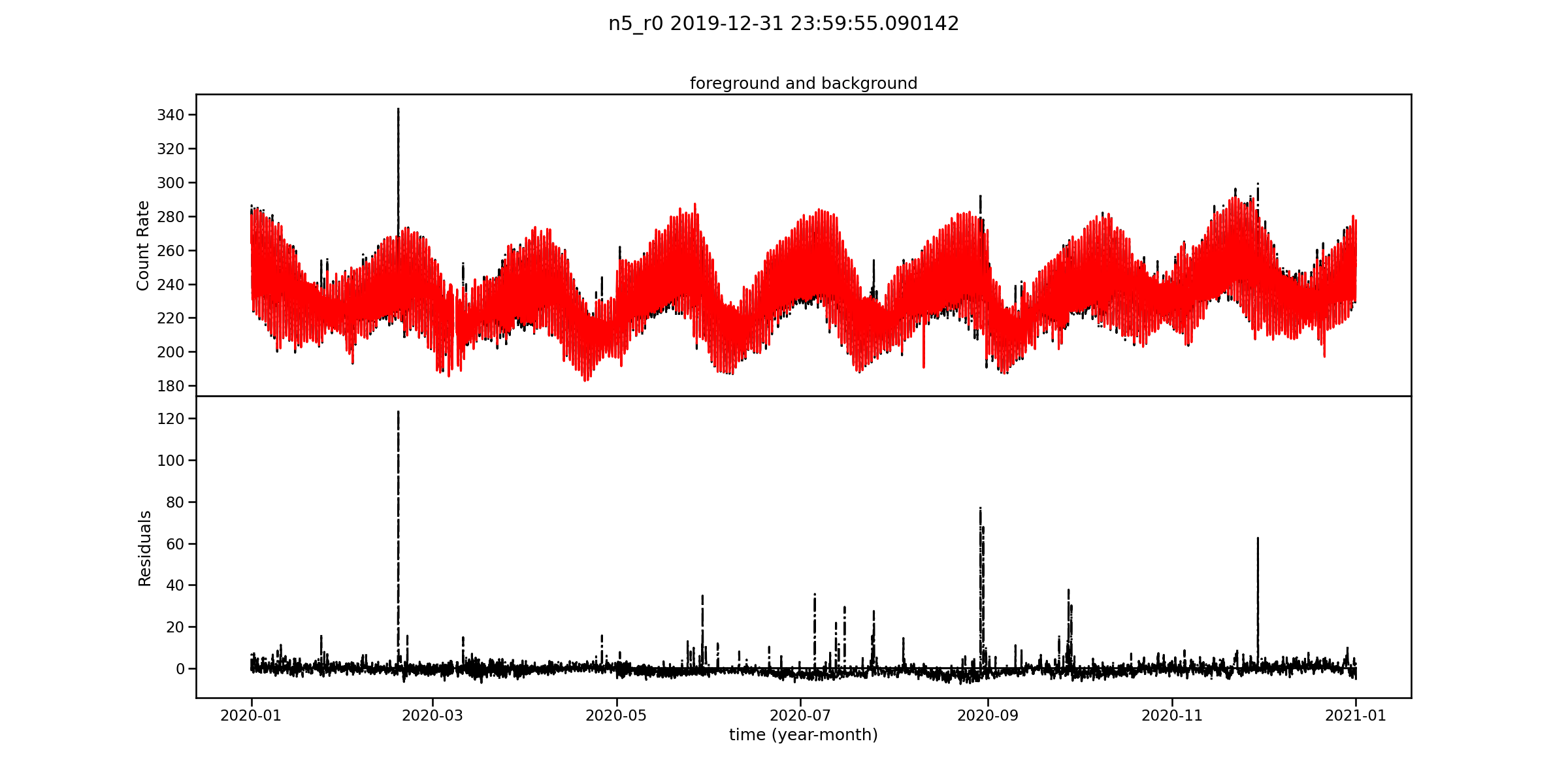}
\caption{\label{fig:solarmin2020_orbit1} The background estimation in year 2020 for detector \texttt{n5} (Sun-facing) in the energy range r0. The count rates with a time bin of 4.096s are the averaged over 1 period orbit (96m).}
\end{figure}

\begin{figure}[H]
\hspace*{-2cm} 
\centering
\includegraphics[width=1.25\textwidth]{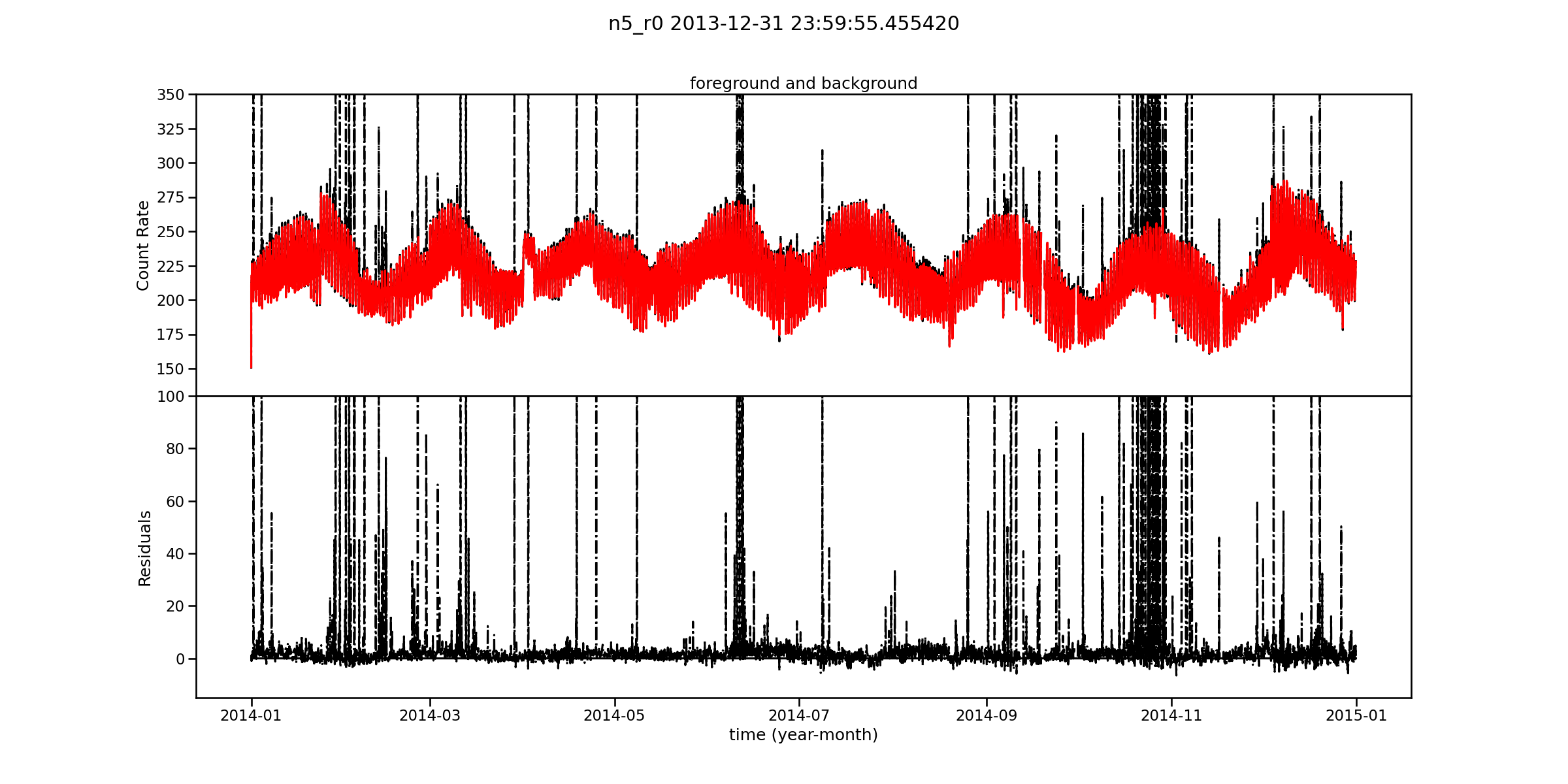}
\caption{\label{fig:solarmaxzoom2014_orbit1} The background estimation in year 2014 for detector \texttt{n5} (Sun-facing) in the energy range r0. The count rates with a time bin of 4.096s are the averaged over 1 period orbit (96m). A zoom-in is applied to avoid the outliers shown in Figure \ref{fig:solarmax2014_orbit1}.}
\end{figure}

\begin{figure}[H]
\hspace*{-2cm} 
\centering
\includegraphics[width=1.25\textwidth]{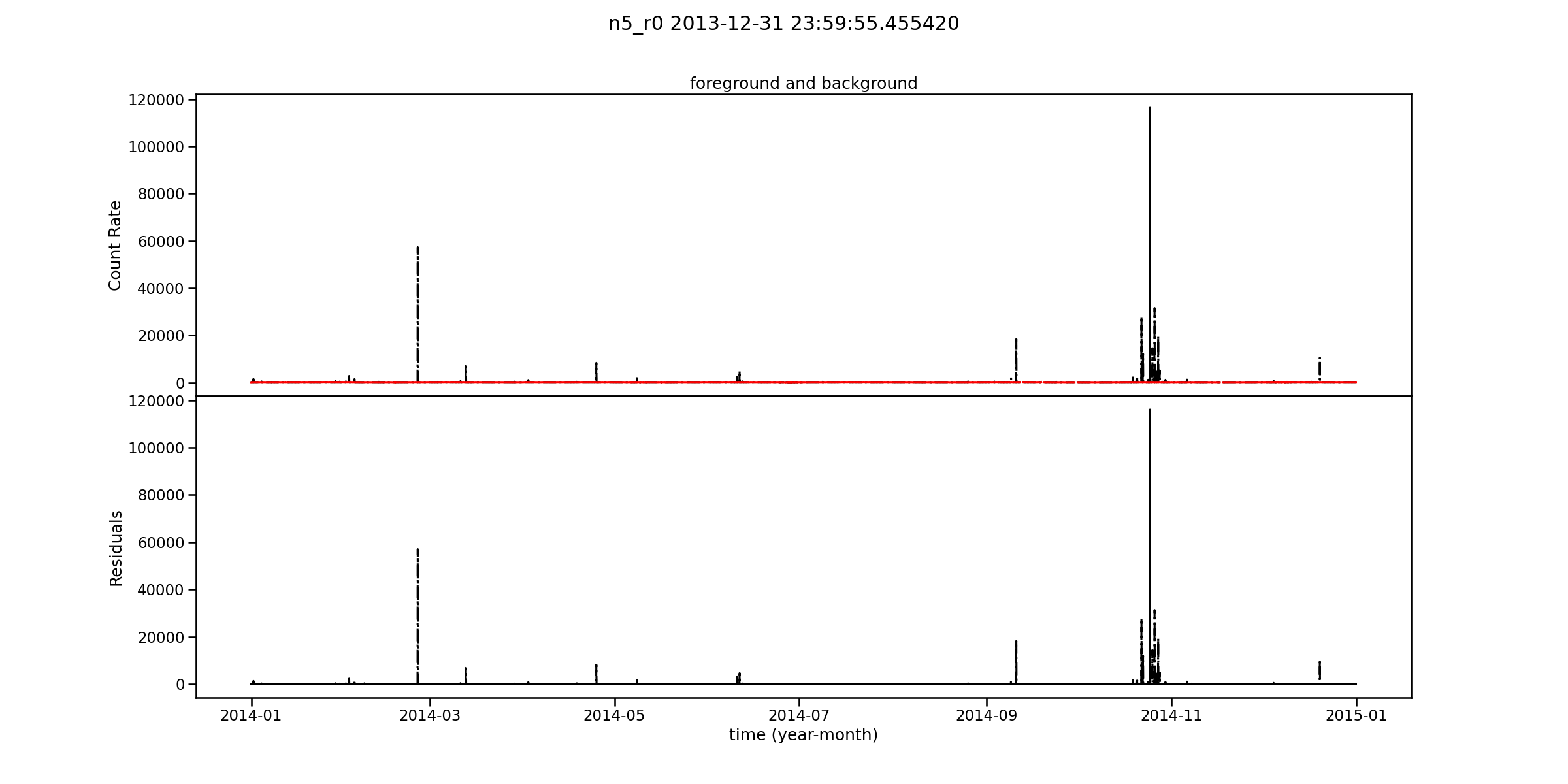}
\caption{\label{fig:solarmax2014_orbit1} The background estimation in year 2014 for detector \texttt{n5} (Sun-facing) in the energy range r0. The solar activity in this year is tremendously high. The count rates with a time bin of 4.096s are the averaged over 1 period orbit (96m).}
\end{figure}

\begin{figure}[H]
\hspace*{-2cm} 
\centering
\includegraphics[width=1.25\textwidth]{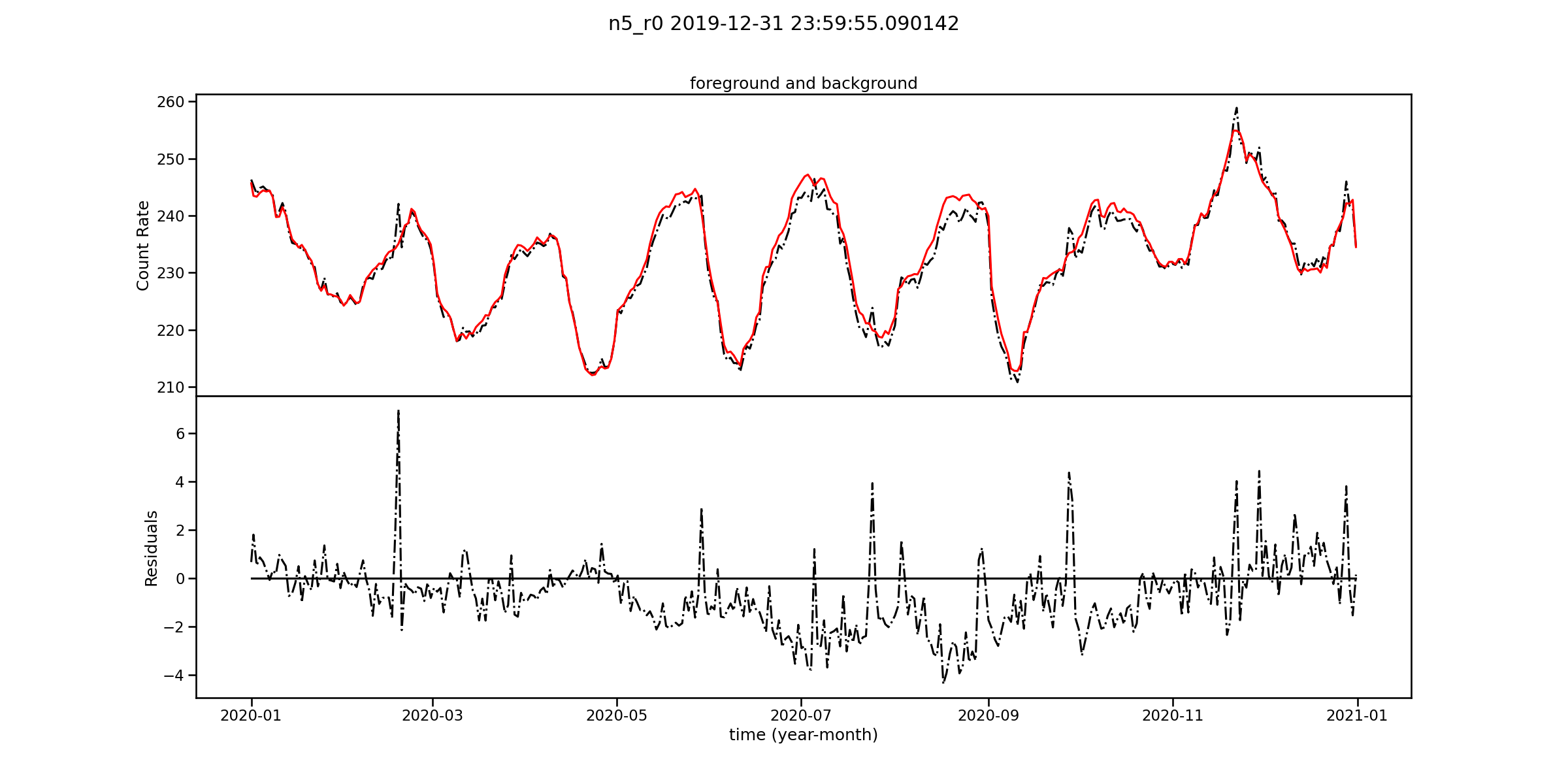}
\caption{\label{fig:solarmin2020_orbit16} The background estimation in year 2020 for detector \texttt{n5} (Sun-facing) in the energy range r0. The count rates with a time bin of 4.096s are the averaged over 16 period orbits ($25.6$h).}
\end{figure}

\begin{figure}[H]
\hspace*{-2cm} 
\centering
\includegraphics[width=1.25\textwidth]{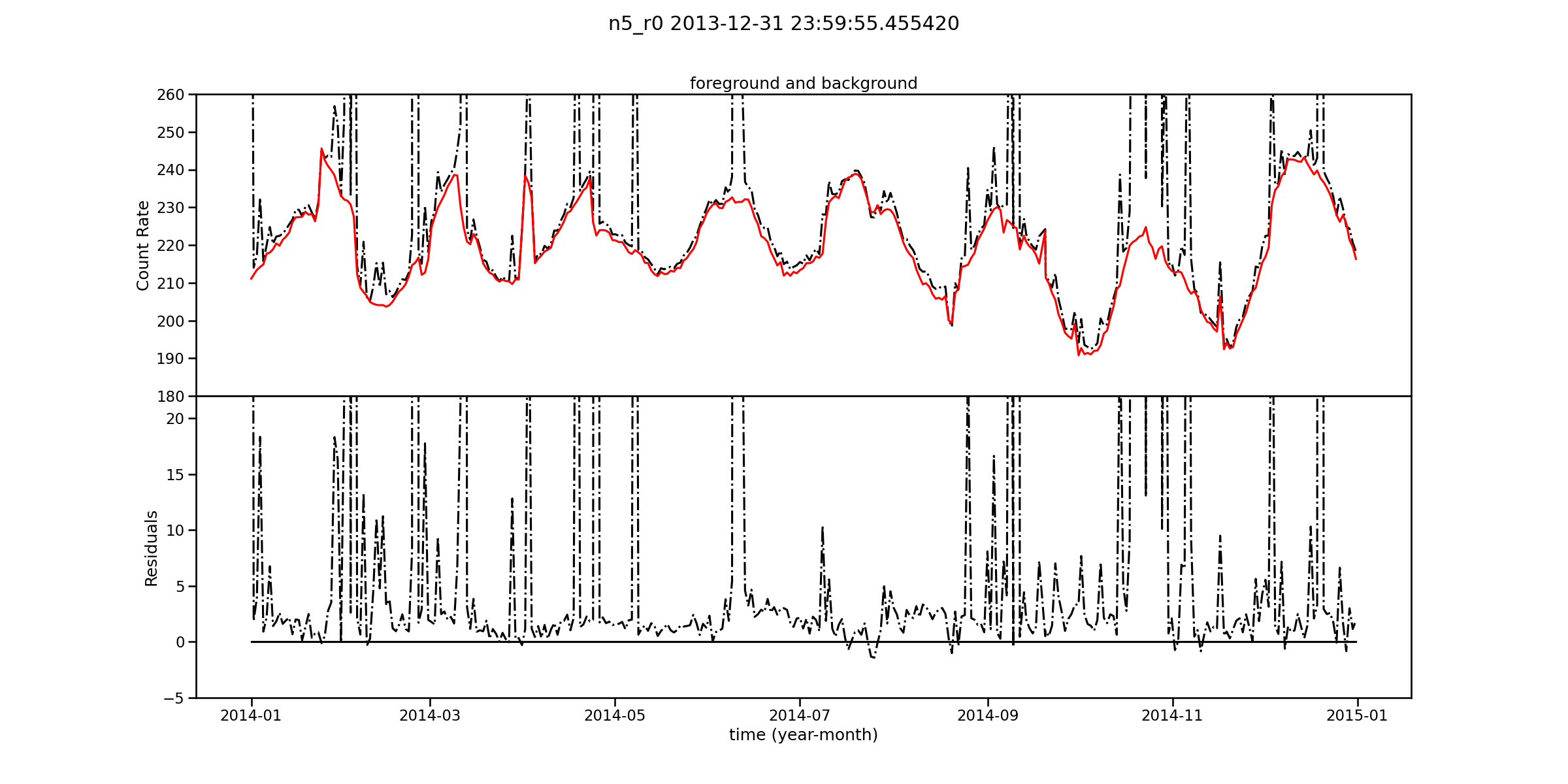}
\caption{\label{fig:solarmaxzoom2014_orbit16} The background estimation in year 2014 for detector \texttt{n5} (Sun-facing) in the energy range r0. The count rates with a time bin of 4.096s are the averaged over 16 period orbits ($25.6$h).}
\end{figure}

Hermes Pathfinder will be launched in 2024 that is near the next solar maximum forecast in 2025 \cite{space2019solar,biesecker2019solar}. This analysis is interesting because reveals what background is expected and how the NN background estimation performs in the two periods. 
The most sensitive detector for the solar activity is the Sun-facing \texttt{n5} \cite{meegan2009fermi}. In this analysis are considered background binned in a GBM period orbits (about 96m) and 16 GBM period orbits, for range 0, the most sensitive for solar flares, in the year of the last solar minima, 2014, and the local minima, 2020. 
The Figures \ref{fig:solarmaxzoom2014_orbit1}, \ref{fig:solarmax2014_orbit1}, \ref{fig:solarmaxzoom2014_orbit16} and
\ref{fig:solarmin2020_orbit1}, \ref{fig:solarmin2020_orbit16} are obtained considering respectively years 2014 and 2020, a NN per each year is trained. One orbit time binning for 2020 \ref{fig:solarmin2020_orbit1} and 2014 \ref{fig:solarmax2014_orbit1} are not comparable due to the high values of the latter but if we zoom the estimated background part we see a comparable counts rate value \ref{fig:solarmaxzoom2014_orbit1}. The same reasoning applies for 16 orbits in 2020 \ref{fig:solarmin2020_orbit16} and 2014 \ref{fig:solarmaxzoom2014_orbit16}. 
In Table \ref{tab:mae_solar} are presented the performance of the background estimation for the year 2014 and 2020. If we use a more robust metric than MAE, such MeAE, we can conclude that the estimations have comparable performance, so the the training procedure of the NN is not influenced by a huge solar activity.

Some reference for the solar cycle prediction can be found in \cite{hathaway1994shape}, \cite{upton2018updated}, \cite{bhowmik2018prediction}. 
For evaluation purposes, the Median Absolute Error (MeAE)
\begin{equation*}
     \text{MeAE}(x, y) = \text{median} ( \{ \mid y_{i} - x_{i} \mid \} )
\end{equation*}
is a robust metric against the outliers.

\begin{table}[!h]
\centering
 \begin{tabular}{||c | c c c c ||} 
 \hline
 year & MAE train & MAE test & MeAE train & MeAE test \\ [0.5ex] 
 \hline\hline
  2014 & 10.601 & 10.467 & 3.940 & 3.944 \\
  2020 & 4.897 & 4.914 & 3.901 & 3.909 \\ [1ex] 
 \hline
 \end{tabular}
 \caption{For each year are presented the metrics averaged per detector and range.}\label{tab:mae_solar}
\end{table}


\section{Interesting events}\label{events}

\begin{figure}[H]
\hspace*{-1cm} 
\centering
\begin{subfigure}{.45\linewidth}
    \centering
    \includegraphics[width=1.25\textwidth]{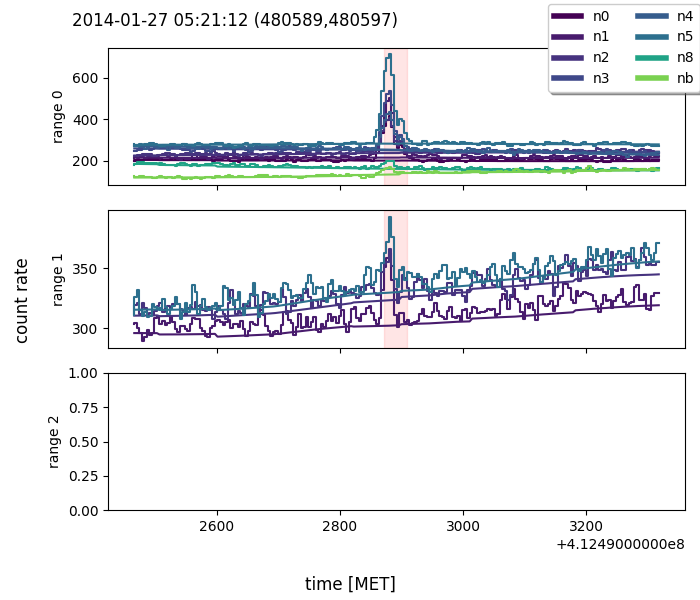}
\end{subfigure}
    \hfill
\begin{subfigure}{.5\linewidth}
    \centering
    \includegraphics[width=1.25\textwidth]{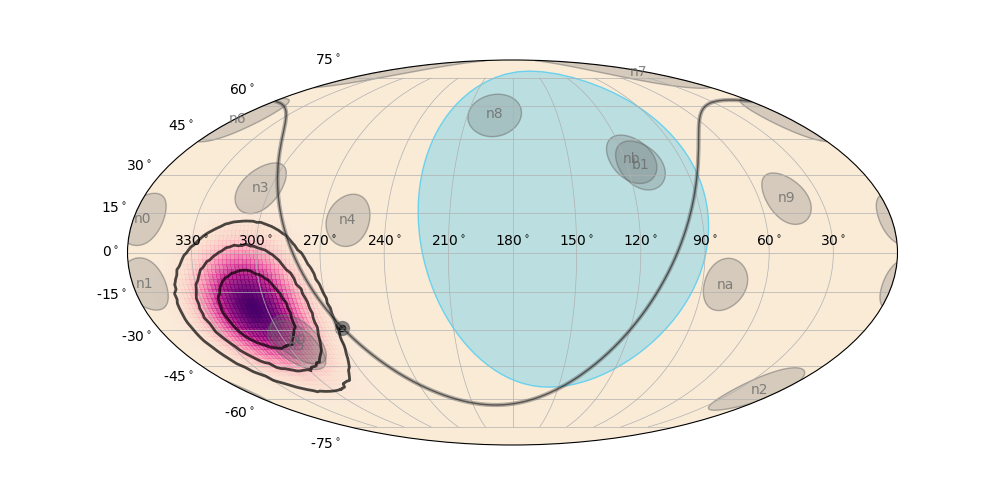}
\end{subfigure}

\caption{Lightcurve and localization for id 1. The Sun is located under the purple spot. We classify this event as SF.}
\label{fig:events1}
\end{figure}

\begin{figure}[H]
\hspace*{-1cm} 
\centering
\begin{subfigure}{.45\linewidth}
    \centering
    \includegraphics[width=1.25\textwidth]{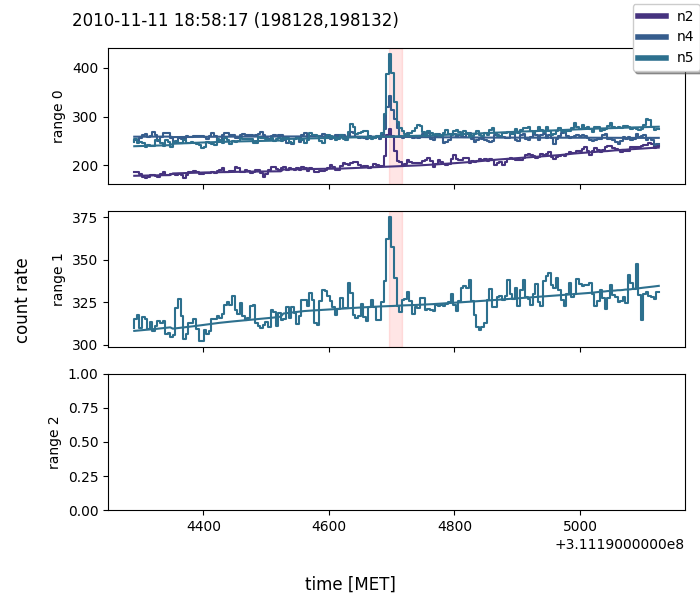}
\end{subfigure}
    \hfill
\begin{subfigure}{.5\linewidth}
    \centering
    \includegraphics[width=1.25\textwidth]{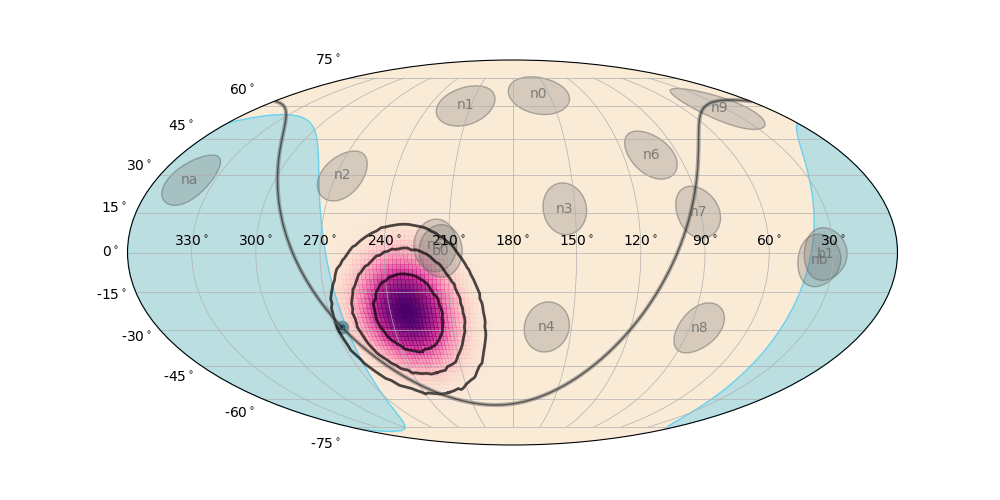}
\end{subfigure}

\caption{Lightcurve and localization for id 2. The Sun is located under the purple spot. We classify this event as SF.}
\label{fig:events2}
\end{figure}

\begin{figure}[H]
\hspace*{-1cm} 
\centering
\begin{subfigure}{.45\linewidth}
    \centering
    \includegraphics[width=1.25\textwidth]{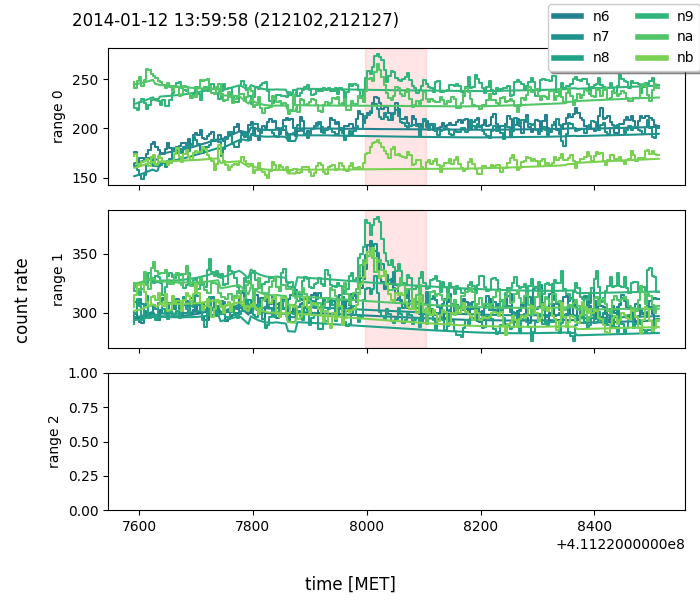}
\end{subfigure}
    \hfill
\begin{subfigure}{.5\linewidth}
    \centering
    \includegraphics[width=1.25\textwidth]{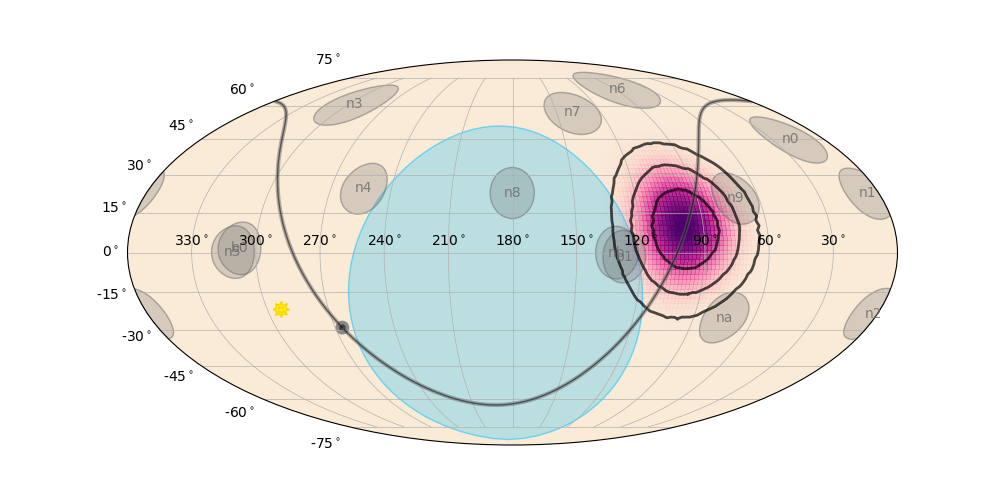}
\end{subfigure}
\caption{Lightcurve and localization for id 3. Because of its location near the galactic plane and the Earth's horizon, we could classify this event as TGF or GF.}
\label{fig:events3}
\end{figure}

\begin{figure}[H]
\hspace*{-1cm} 
\centering
\begin{subfigure}{.45\linewidth}
    \centering
    \includegraphics[width=1.25\textwidth]{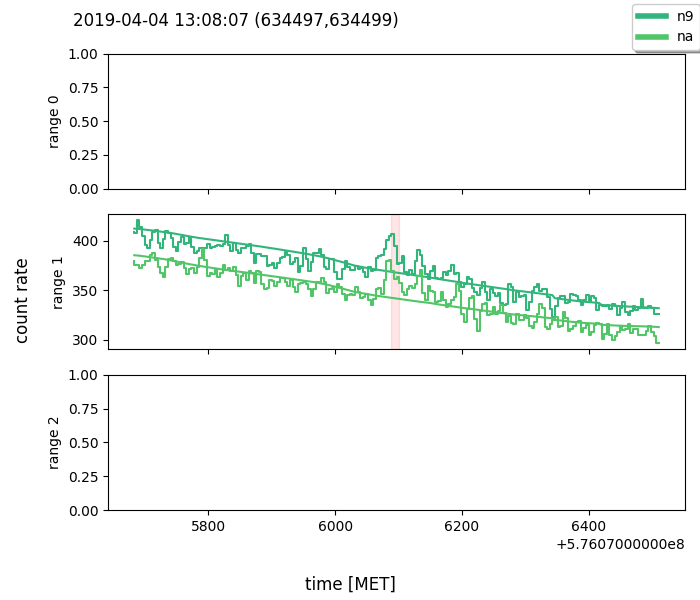}
\end{subfigure}
    \hfill
\begin{subfigure}{.5\linewidth}
    \centering
    \includegraphics[width=1.25\textwidth]{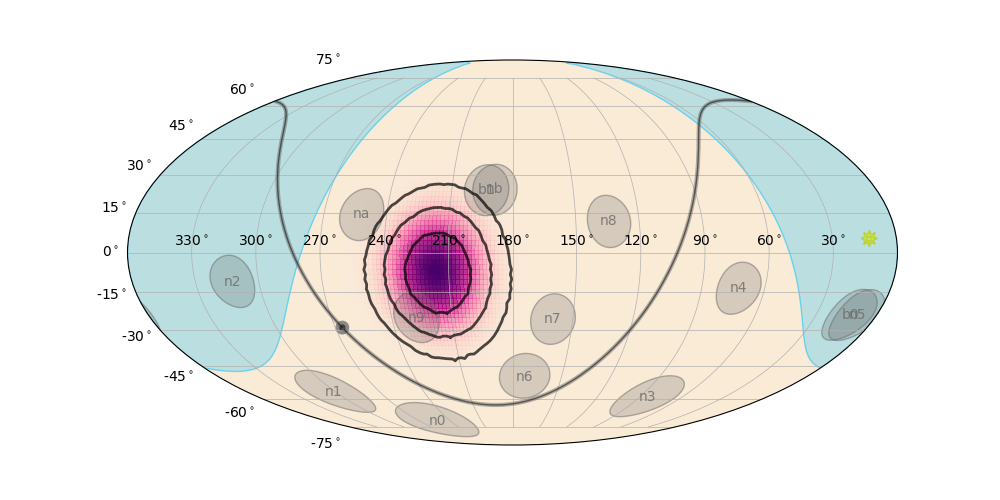}
\end{subfigure}
\caption{Lightcurve and localization for id 4. We classify this event as a GRB.}
\label{fig:events4}
\end{figure}

\begin{figure}[H]
\hspace*{-1cm} 
\centering
\begin{subfigure}{.45\linewidth}
    \centering
    \includegraphics[width=1.25\textwidth]{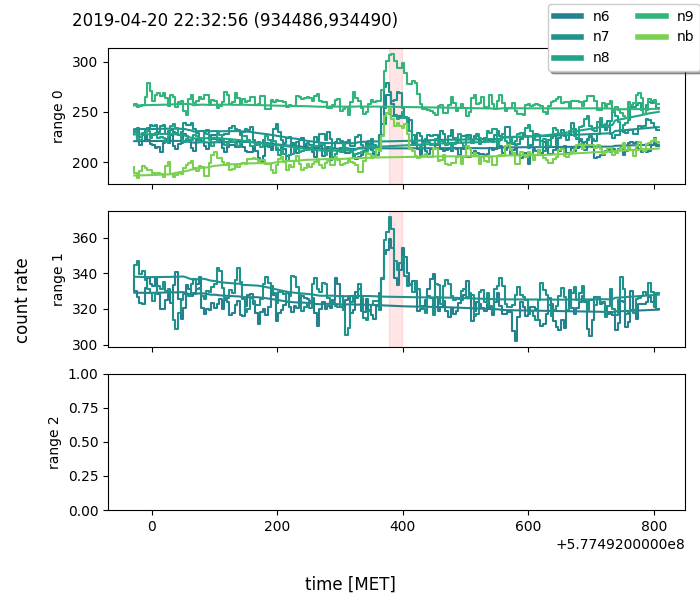}
\end{subfigure}
    \hfill
\begin{subfigure}{.5\linewidth}
    \centering
    \includegraphics[width=1.25\textwidth]{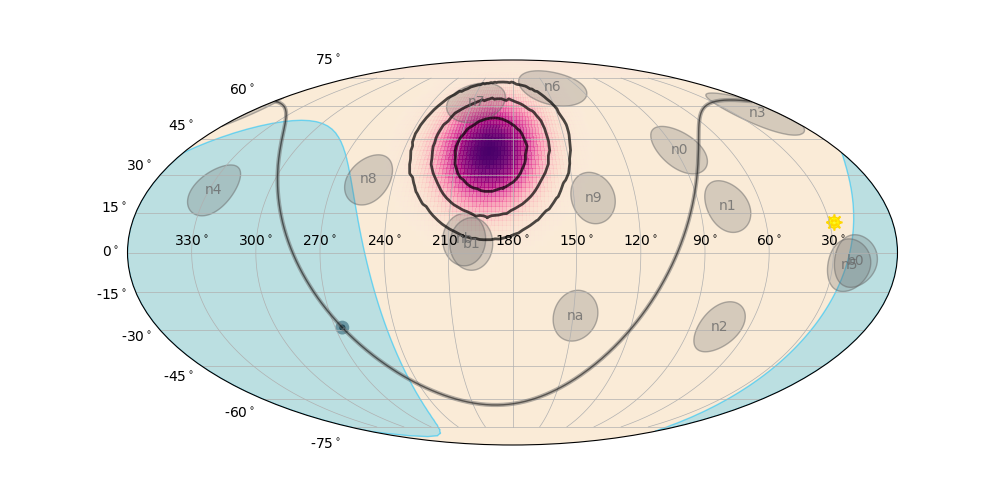}
\end{subfigure}
\caption{Lightcurve and localization for id 5. We classify this event as a GRB.}
\label{fig:events5}
\end{figure}

\begin{figure}[H]
\hspace*{-1cm} 
\centering
\begin{subfigure}{.45\linewidth}
    \centering
    \includegraphics[width=1.25\textwidth]{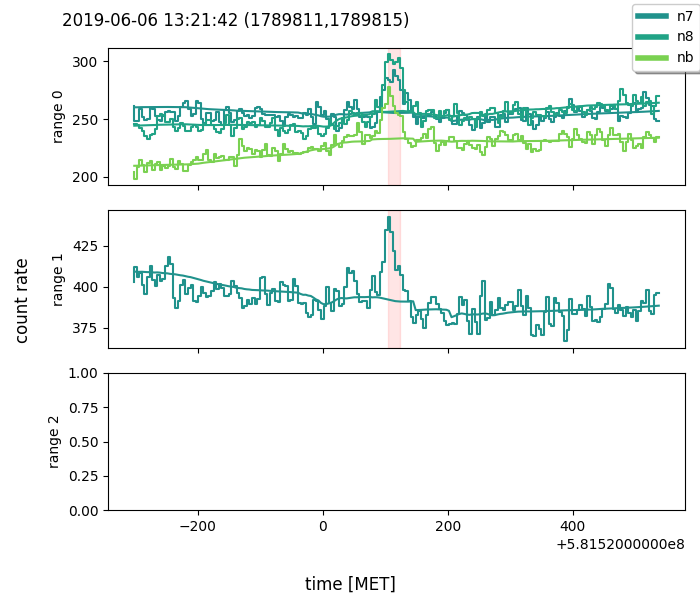}
\end{subfigure}
    \hfill
\begin{subfigure}{.5\linewidth}
    \centering
    \includegraphics[width=1.25\textwidth]{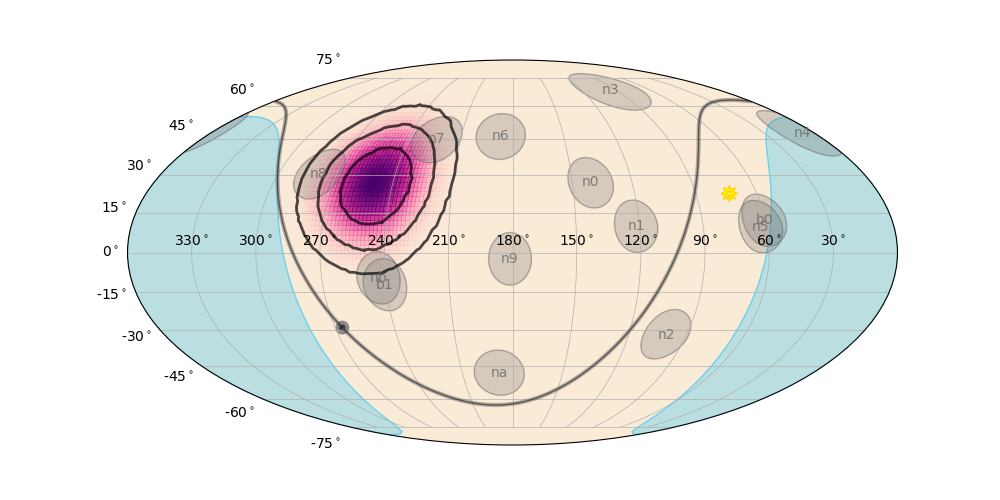}
\end{subfigure}
\caption{Lightcurve and localization for id 6. We could classify this event as a GRB or, because of its proximity to the galactic plane, a GF.}
\label{fig:events6}
\end{figure}

\begin{figure}[H]
\hspace*{-1cm} 
\centering
\begin{subfigure}{.45\linewidth}
    \centering
    \includegraphics[width=1.25\textwidth]{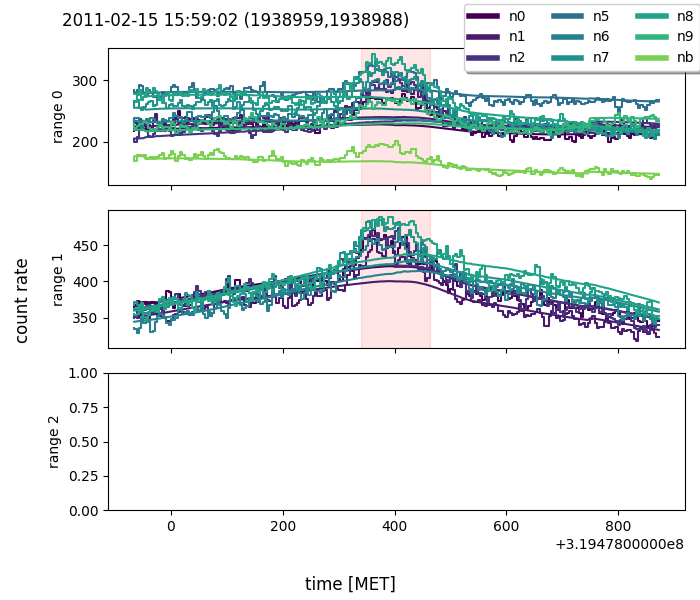}
\end{subfigure}
    \hfill
\begin{subfigure}{.5\linewidth}
    \centering
    \includegraphics[width=1.25\textwidth]{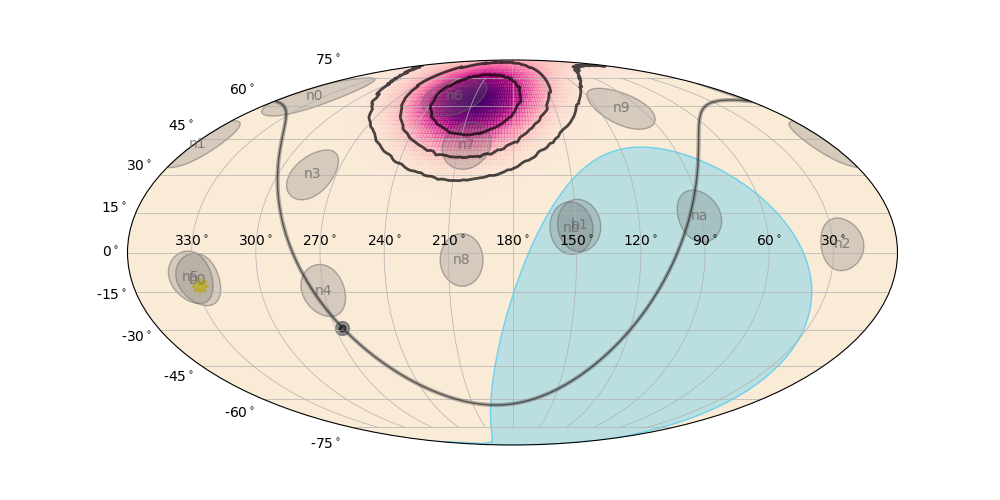}
\end{subfigure}
    \hfill
\begin{subfigure}{1.\linewidth}
    \centering
    \includegraphics[width=1.\textwidth]{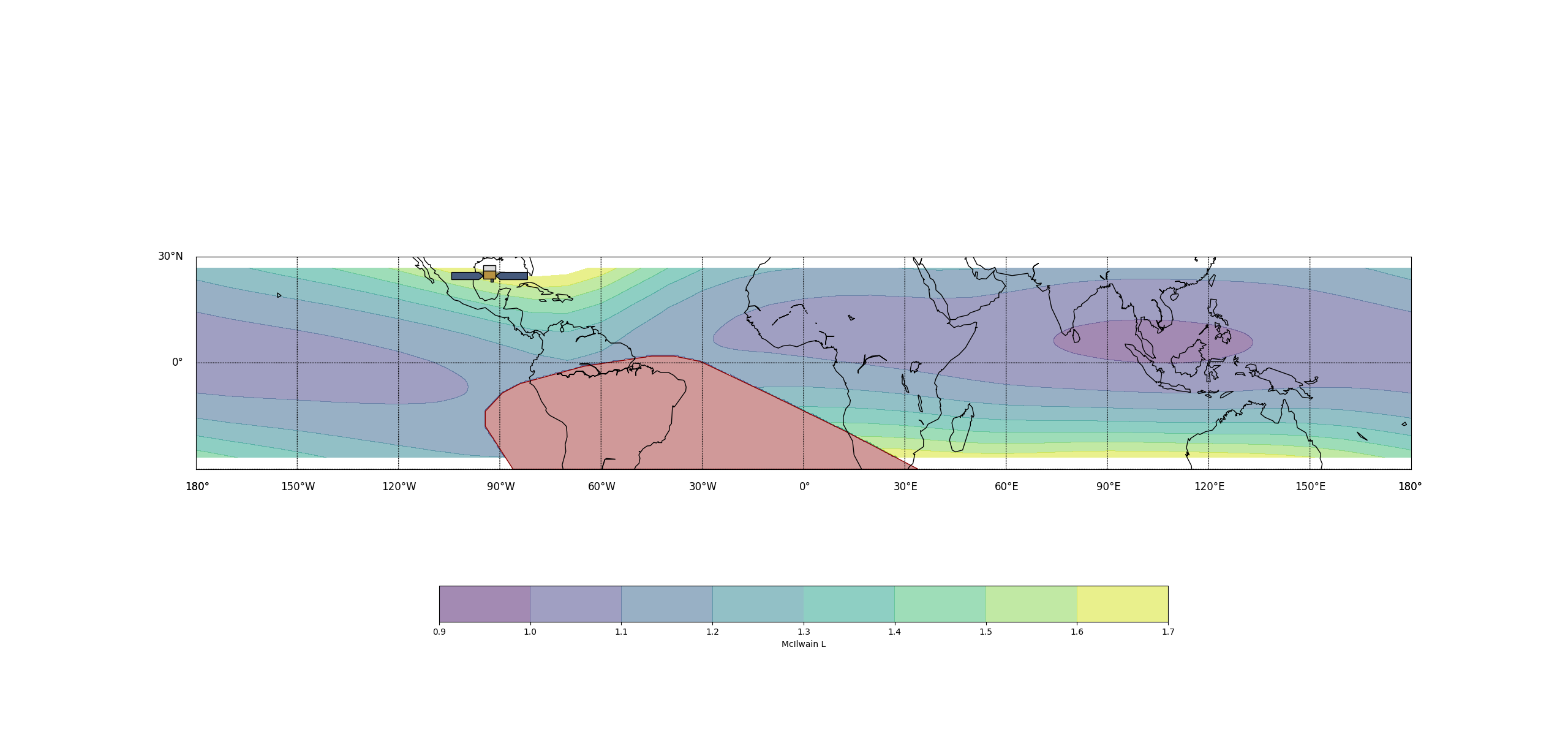}
\end{subfigure}
\caption{In the first two figures, the lightcurve and localization for id 7 are shown. The third one shows where the GBM satellite is located on Earth. Local Particles events like LOCLPAR1905205 and LOCLPAR1904085 have occurred in this region. We classify this event as uncertain.
\label{fig:events7}}
\end{figure}

\newpage
\section{Catalog tables}
\begin{small}
\begin{longtable}{llp{0.1\textwidth}rp{0.1\textwidth}p{0.15\textwidth}llll}
\toprule
{\#} & ID & Trigger time &  T (s) & Detectors triggered & Catalog trigger name & S r0 & S r1 & S r2 & CE \\
\midrule
\endhead
1   &    2010\_1 &  2010-11-02 14:16:36 &      8.19 &                                   n2 &       UNKNOWN: GRB/GF &     3.99 &     3.38 &        0 &      P \\
6   &    2010\_6 &  2010-11-11 13:04:23 &     32.77 &                                   n2 &          UNKNOWN: GRB &     4.29 &     5.81 &        0 &      P \\
7   &   2010\_7* &  2010-11-11 18:58:17 &     16.38 &                             n2 n4 n5 &           UNKNOWN: SF &    $>10$ &     4.57 &        0 &      R \\
10  &   2010\_10 &  2010-11-12 23:46:52 &     19.73 &                                   n1 &       UNKNOWN: GRB/GF &        0 &     9.18 &     5.31 &      P \\
14  &   2010\_14 &  2010-11-22 13:23:34 &      8.19 &                                n5 nb &  UNKNOWN: UNC(LP)/GRB &        0 &     7.45 &     5.41 &      R \\
33  &   2010\_34 &  2010-12-28 18:22:27 &    106.50 &                                   na &  UNKNOWN: TGF &        0 &     8.08 &        0 &      P \\
38  &   2010\_40 &  2011-01-16 11:40:21 &     20.48 &                                n2 n5 &          UNKNOWN: GRB &        0 &     5.36 &        0 &      S \\
44  &   2010\_50 &  2011-01-31 15:52:43 &      8.19 &                             n5 n6 nb &      UNKNOWN: UNC(LP) &        0 &     7.08 &     4.27 &      R \\
\end{longtable}

\begin{longtable}{llp{0.1\textwidth}rp{0.1\textwidth}p{0.15\textwidth}llll}
\toprule
{\#} & ID & Trigger time &  T (s) & Detectors triggered & Catalog trigger name & S r0 & S r1 & S r2 & CE \\
\midrule
\endhead
0   &    2010\_0 &  2010-11-01 21:34:05 &     34.81 &                             n3 n4 n5 &     GRB101101899 &     5.54 &     8.51 &        0 &      R \\
2   &    2010\_2 &  2010-11-02 20:10:09 &     41.47 &                       n6 n7 n8 n9 nb &     GRB101102840 &      5.9 &    $>10$ &        0 &      R \\
3   &    2010\_3 &  2010-11-03 12:13:28 &    114.69 &           n0 n1 n2 n3 n4 n5 n9 na nb &   SFLARE10110350 &    $>10$ &    $>10$ &        0 &      R \\
4   &    2010\_4 &  2010-11-04 19:26:16 &      4.43 &                                   n2 &     GRB101104810 &        0 &     3.22 &        0 &      P \\
5   &    2010\_5 &  2010-11-07 00:16:24 &    341.51 &                       n0 n1 n3 n4 n5 &     GRB101107011 &     8.84 &    $>10$ &        0 &      R \\
8   &    2010\_8 &  2010-11-12 22:10:31 &     17.92 &                                n2 n5 &     GRB101112924 &     5.36 &     7.77 &        0 &      R \\
9   &    2010\_9 &  2010-11-12 23:36:52 &     96.26 &                          n3 n6 n7 n8 &     GRB101112984 &    $>10$ &    $>10$ &        0 &      R \\
11  &   2010\_11 &  2010-11-13 11:35:33 &    202.24 &           n0 n1 n2 n6 n7 n8 n9 na nb &     GRB101113483 &    $>10$ &    $>10$ &        0 &      R \\
\end{longtable}
\end{small}
\clearpage

\end{document}